\documentclass[twocolumn,prb,aps,amsfonts]{revtex4-1}
\usepackage{latexsym}
\usepackage{dcolumn}
\usepackage[dvips]{graphicx}
\usepackage{amssymb}
\usepackage{graphicx}
\usepackage{psfrag, color}
\usepackage{verbatim}
\usepackage{amsmath}
\usepackage{epsf}

\begin{document}

\draft

\title{Mobility versus quality in 2D semiconductor structures} 
\author{S. Das Sarma$^1$ and E. H. Hwang$^{1,2}$}
\address{$^1$Condensed Matter Theory Center, 
Department of Physics, University of Maryland, College Park,
Maryland  20742-4111 \\
$^2$SKKU Advanced Institute of Nanotechnology and Department of Physics, Sungkyunkwan
University, Suwon, 440-746, Korea} 
\date{\today}

\begin{abstract}
We consider theoretically effects of random charged impurity disorder
on the {\it quality} of high-mobility two dimensional (2D)
semiconductor structures, 
explicitly demonstrating that the sample mobility is not necessarily a
reliable or universal indicator of the sample quality in
high-mobility modulation-doped 2D GaAs structures because, depending
on the specific system property of interest, mobility and quality
may be controlled by different aspects of the underlying disorder
distribution, particularly since these systems are dominated by
long-range Coulomb disorder from both near and far random quenched
charged impurities. 
We show that in the presence of both channel and remote
charged impurity scattering, which is a generic situation in
modulation-doped high-mobility 2D carrier systems, it is quite
possible for higher (lower) mobility structures to have lower (higher)
quality as measured by the disorder-induced 
single-particle level broadening. In particular, we
establish that there is no reason to expect a unique relationship
between mobility and quality in 2D semiconductor structures as both
are independent functionals of the disorder distribution, and are
therefore, in principle, independent of each other. Using a simple,
but reasonably realistic, ``2-impurity'' minimal model of the disorder
distribution, we provide concrete examples of situations where higher
(lower) mobilities correspond to lower (higher) sample qualities. We
discuss experimental implications of our theoretical results and
comment on possible strategies for future improvement of 2D sample quality.
\end{abstract}

\maketitle

\section{introduction}

One of the most significant materials developments in the fundamental
quantum condensed matter physics, which is not universally known
outside the 2D community, has been the astonishing 3,000-fold increase
in the low temperature electron mobility of GaAs-based 2D confined
quantum systems from $ \sim 10^4$ cm$^2$/Vs in the first modulation-doped
GaAs-AlGaAs 2D heterostructures \cite{stormer} in 1978 to the current
world-record mobility of $\sim 3-4 \times 10^7$ cm$^2$/Vs  in the
best available modulation doped GaAs-AlGaAs quantum wells of today.
\cite{heiblum}
This represents a truly remarkable more than three orders of magnitude
enhancement in the low temperature ($\sim 1$K) electron mean free path
from a rather short microscopic length $\sim 50$ nm in 1978 to the
essentially macroscopic length scale of $\sim 0.2$ mm in 2010. This
incredible 3,000-fold enhancement of the 2D carrier mean free path,
although much less well-known than the celebrated Moore's law in the
Si electronics industry, is actually quantitatively on par with
Moore's law increase in the microprocessor performance. Unlike Moore's
law in Si microelectronics performance, where the motivation has been
technological, however, the drive for the mobility enhancement in 2D
GaAs structures has been motivated entirely by fundamental physics
considerations. Indeed, this increase in the 2D mobility has been
accompanied by some of the most spectacular experimental discoveries
in modern physics including, for example, the fractional quantum Hall
effect (FQHE) \cite{tsuiprl1982}, the even-denominator FQHE
\cite{willettprl1987},  
the bilayer half-filled FQHE
\cite{shayeganprl1992}, the anisotropic stripe and
bubble phases \cite{lillyprl}, and many other well-known novel 2D
phenomena. (As an aside we point out that, qualitatively similar to
the situation in the Moore's law, the exponential enhancement in the
2D mobility of semiconductor heterostructures has slowed down
considerably in the recent years with only a 30\% increase in the
mobility during the $2003-2013$ period, from $\sim 3\times 10^7$ to
$\sim 4 \times 10^7$ cm$^2$/Vs, after roughly a factor of 1,000
increases during $1978-2003$.)

We mention right in the beginning that our interest here is obviously
the $T=0$ transport properties (or low-temperature transport properties)
with the temperature being much smaller than both the Fermi
temperature and the Bloch-Gr\"{u}neisen (BG) temperature of the 2D system so
that all thermal effects have saturated, and we do not need to account
for either phonon scattering or finite temperature effects in the
Fermi distribution function.  The high mobility of 2D semiconductor
structures applies only to this low-temperature situation, and at
higher temperatures ($>10$K), the mobility is dominated by phonon
scattering, a situation already well-studied in the literature\cite{stormerbg}.
Our theory is just restricted only to $T=0$ impurity-scattering-limited 2D
transport properties, which limit the ultimate achievable mobility in
these systems.  It is also important to emphasize that we ignore all
weak localization aspects of 2D transport properties, restricting
entirely to the semiclassical transport behavior where the concept of
a mobility is valid.  Thus, the theory applies only at densities where
weak-localization behavior does not manifest itself at the
experimental temperatures ($\sim 25$ mK $- 2$ K).  At a fixed density, this
limits our theory to a temperature high enough ($>1$ mK) so that the
phase breaking length is shorter than the semiclassical mean free
path.  The physics of mobility/quality dichotomy discussed in this
paper satisfies all of these constraints (i.e. no phonon scattering,
no weak localization, temperature much lower than the fermi
temperature) very well.  

Although the {\it mobility} enhancement of 2D systems has generally
led to the improvement of sample {\it quality} on the average over the
years as manifested in the observation of new phenomena, it has been
known from the early days that the connection between {\it mobility}
and {\it quality} is at best a statistically averaged statement over
many samples and is not unique, i.e., a sample with higher mobility
than another sample may not necessarily have a higher quality with
respect to some specific property (e.g., the existence or not of a
particularly fragile fractional quantum Hall plateau). Thus, higher
(lower) mobility does not necessarily always translate into higher
(lower) quality for specific electronic properties. Since 2D carrier
mobility is a function of carrier density \cite{dassarmaprb2013}, in fact, it is, in
principle, possible for a sample to have a higher (lower) mobility
than another sample at higher (lower) carrier density, implying that
the measured mobility at some fixed high density is not (always) even
a good indicator of transport quality itself as a function of carrier
density, let alone being an indicator for the quality of other
electronic properties!

The reason for the above-mentioned mobility/quality dichotomy is
rather obvious to state, but not easy to quantify. Both mobility and
quality (e.g., the measured activation gap for a specific FQHE state
or some other specified electronic property) depend on the full
disorder distribution affecting the system which is in general both
unknown and complex, and depends also on the sample carrier density in
a complicated manner. 
The disorder distribution is characterized by
many independent parameters, and therefore, all physical properties of
the system, being unique functionals of the disorder distribution, are
independent of each other. In particular, the dc conductivity
$\sigma$, which determines the density-dependent mobility 
$\mu = \sigma/ne$ where $n$ is the 2D carrier density (and $e$ the
magnitude of electron charge), is determined by essentially an
integral over the second moment of the disorder distribution whereas
other physical properties (i.e., the quality, although there could, in
principle, be many independent definitions of sample quality depending
on independent experimental measurements of interest) could be
determined by other functionals of the disorder distribution. Thus,
for any realistic disorder distribution, we do not expect any unique
relationship between mobility and quality, and it should be possible,
in principle, for samples of different mobility to have the same
quality or vice versa (i.e., samples of different quality to have the
same 2D mobility). 

There is still the vague qualitative expectation, however, that if the
sample mobility is enhanced by improving the sample purity (i.e.,
suppressing disorder), then this should automatically also improve the
sample quality (perhaps not necessarily by the same quantitative
factor) since reduced disorder should enhance quality. We will show
below that this may not always be the case since mobility and quality
(for a specific property) may be sensitive to completely different
aspects of disorder and therefore enhancing mobility by itself may do
nothing to improve quality in some situations. On the other hand, when
mobility and quality are both determined by exactly the same
microscopic disorder in the sample, increasing (decreasing) one would
necessarily improve (suppress) the other
although we will see later in this work [see Fig.~\ref{fig01}(c)], for
example) this is not necessarily true if mobility and quality are both
determined by remote dopant scattering arising from long-range Coulomb
disorder. 

It is important to emphasize a subtle aspect of the mobility/quality
dichotomy with respect to experimental samples. Theoretically, we can
consider a hypothetical system with continuously variable disorder
(i.e., the parameters characterizing the disorder distribution such as
the quenched charged impurity density and their strength as well as
their spatial locations including possible spatial correlations in the
impurity distribution can all be varied at will). Experimentally,
however, the situation is qualitatively different. One does not
typically change the impurity distribution in a sample in a controlled
manner and make measurements as a function of disorder
distribution. Experimentally, measurements are made in {\it different}
samples and compared, and in such a situation there is no reason to
expect two samples with identical mobility at some specified carrier
density to have identical impurity distributions. The impurity
distributions in different samples can be considered to be identical
only if the full density-dependent mobility $\mu(n)$, or equivalently
the density-dependent conductivity $\sigma(n)$, are identical in all
the samples. Such a situation of course never happens in practice, and
typically when experimental mobilities in different samples are quoted
to be similar in magnitudes, one is referring to either the maximum
mobility (occurring typically at different carrier densities in
different samples) or the mobility at some fixed high carrier density
(and not over a whole range of carrier density). If two samples happen
to have the same maximum mobility or the same mobility at one fixed
density, there is no reason to expect them to have the same quality
with respect to all experimental properties at arbitrary
densities. This obvious aspect of mobility versus 
quality dichotomy has not much been emphasized in the literature.

It should be clear from the discussion above that in the `trivial'
(and experimentally unrealistic in 2D semiconductor structures)
situation of the system having just one type of impurities uniquely
defining the applicable impurity distribution, both mobility and
quality, by definition, would be determined by exactly the same
impurity configuration since there is just one set of impurities by
construction. Such a situation is, in fact, common in 3D
semiconductors where the applicable disorder is almost always
described by a random uniform background of uncorrelated quenched
charged impurity centers, which can be uniquely characterized by a
single 3D impurity density $n_i$. Obviously, in this situation both
mobility and quality are uniquely defined by $n_i$, and thus
increasing (decreasing) $n_i$ would decrease (increase) both mobility
and quality (although not necessarily by the same quantitative factor
since mobility and quality are likely to be different functions of
system parameters in general). In such a simplistic situation,
mobility and quality are likely to monotonically connected by a unique
relationship, and hence enhancing system mobility should always
improve the system quality since the same disorder determines both
properties. 
In fact, this simple situation always applies if the mobility/quality
are both limited by purely short-range disorder in the system.

By contrast, 2D semiconductor structures almost always have several
qualitatively distinct disorder mechanisms affecting transport and
other properties arising from completely different physical
origins. For example, it is well-known \cite{andormp} that Si-MOSFETs
have at least three different operational disorder mechanisms: random
charged impurities in the insulating SiO$_2$ oxide layer, in the bulk
Si itself, and random short-range surface roughness at the Si-SiO$_2$
interface. There may still be other distinct scattering mechanisms
associated with still different disorder sources in Si-MOSFETs such as
neutral defects or impurities, making the whole situation quite
complex. In MOSFETs, low-density carrier transport is controlled by
long-range charged impurity scattering whereas the high-density
carrier transport is controlled by short-range surface roughness
scattering, and this dichotomy may very well lead to situations where
a measured electronic property (i.e., ``quality'') does not
necessarily correlate with the high-density maximum mobility of the
system. In high-mobility 2D GaAs-AlGaAs-based systems of interest in
the current work, there are at least six distinct scattering
mechanisms of varying importance arising from different physical
sources of disorder in the system. These are: Unintentional background
charged impurities in the 2D GaAs conducting layer; remote dopant
impurities in the insulating AlGaAs layer (which are necessary for
introducing 2D carriers to form the 2DEG); short-range interface
roughness at the GaAs-AlGaAs interfaces; short-range disorder in the
insulator AlGaAs layer arising from alloy disorder (and neutral
defects); unintentional background charged impurities in the
insulating AlGaAs barrier regime; random charged impurities at the
GaAs-AlGaAs interface. This is obviously a highly complex situation
where the complete disorder distribution will have many independent
parameters, and in general, there is no reason to expect a unique
relationship between mobility and quality since mobility could be
dominated by one type of disorder (e.g., background unintentional
charged impurities) and quality may be dominated by a different type
of disorder (e.g., remote dopants far from the 2D layer).

It is clear from the above discussion that the minimal disorder model
capable of capturing the mobility versus quality dichotomy in
high-quality 2D semiconductor structures is a ``2-impurity'' model
with one type of impurity right in the 2D layer itself (arising, for
example, from the unintentional background impurities in the system)
and the other type of impurity being a remote layer separated by a
distance `$d_R$' from the 2D electron layer. This minimal 2-impurity
model is characterized by three independent parameters: $n_R$ and
$d_R$, denoting respectively the 2D charged impurity density in the
remote dopant layer separated by a distance $d_R$ from the 2D
carriers, and $n_B$, the 2D impurity density corresponding to the
unintentional background impurities in the 2D layer with $d=0$. It
is easy to go beyond this minimal model and consider the remote
impurities to be distributed over a finite distance (rather than
simply being placed in a $\delta$-function like layer located at a
distance $d_R$ from the 2D system) or assume the background impurities
to be distributed three-dimensionally (rather than in a 2D plane at
$d=0$), but such extensions do not modify any of our qualitative
conclusion in the current paper as we have explicitly checked
numerically. Also, we discuss our theory assuming a strict 2D model
(with zero thickness) for the electron layer because the finite quasi-2D
layer thickness has no qualitative effect on the physics of
quality versus mobility being discussed in this work. Many of our
numerical results are, however, obtained by incorporating the
appropriate quantitative effects of the quasi-2D layer thickness of
the 2D carriers in calculating the mobility and the quality of the
system within the 2-impurity model.

One last issue we need to discuss in the Introduction is the question
of how to define the sample quality since it is obviously not a unique
property and depends on the specific
experiment being carried out. In order to keep things both simple and
universal, we have decided to use the level broadening or the Dingle
temperature as a measure of the sample quality. The level broadening
$\Gamma$ is defined as $\Gamma = \hbar/2\tau_q$, where $\tau_q$ is
the quantum scattering time (or the single particle relaxation time)
in contrast to the mobility scattering time (or the transport
relaxation time) $\tau_t$ which defines the conductivity ($\sigma$) or
the carrier mobility ($\mu$) through $\sigma = ne^2\tau_t/m$ or $\mu =
ne \tau_t$. In general, $\tau_q \leq \tau_t$ with the equality holding for
purely short-range disorder scattering. Thus, for pure $s$-wave
$\delta$-function short-range scattering model, mobility and quality
are identical for obvious reasons. For a strict 1-impurity model of
underlying sample disorder, mobility ($\tau_t$) and quality ($\tau_q$)
are both affected by the same impurity density, and hence they must
behave monotonically (but not necessarily identically) with changing
disorder, i.e., if the impurity density is decreased (increased) at a
fixed carrier density, both
mobility and quality must increase (decrease) as well.

As emphasized already, however, the 2-impurity model (near and far
impurities or background and remote impurities) introduces a new
element of physics by allowing for the possibility that mobility and
quality could possibly be affected more strongly by different types of
disorder, for example, mobility (quality) could be dominated by near
(far) impurities in the 2-impurity disorder model, thus allowing, in
principle, the possibility of mobility and quality being completely
independent physical properties of realistic 2D semiconductor samples
at least in some situations. We find this situation to be quite
prevalent in very high-mobility 2D semiconductor structures where the
mobility (quality) seems to be predominately determined by near (far)
impurities. In low mobility samples, on the other hand, the situation
is simpler and both mobility and quality are typically determined by
the same set of impurities (usually the charged impurities close to
the 2D layer itself).

In section II we describe our model giving the theoretical formalism
and equations for the 2-impurity model. In section III we provide
detailed results for mobility and quality along with discussion. We
conclude in section IV with a summary of our findings along with a
discussion of our approximations and of the open questions.

\section{model and theory}

We assume a 2D electron (or hole) system at $T=0$ located at the $z=0$
plane with the 2D layer being in the x-y plane in our notation. The 2D
system is characterized entirely by a carrier effective mass ($m$)
defining the single-particle kinetic energy ($E_{\bf k}=\hbar^2k^2/2m$ with
{\bf k} as the 2D wave vector), a background lattice dielectric
constant ($\kappa$) defining the 2D Coulomb interaction [$V({\bf q}) = 2\pi
e^2/\kappa q$ with {\bf q} as the 2D wave vector), and a 2D carrier
density ($n$).

As discussed in the Introduction, we use a minimal 2-impurity model
for the static disorder in the system characterized by three
independent parameters: $n_R$, $d_R$, $n_B$. Here $n_R$ ($n_B$) is the
effective 2D charged impurity density for the remote (background)
impurities with the remote (background) impurities being distributed
randomly in the 2D x-y plane at a distance `$d$' from the 2D
electron system in the $z$-direction with $d=d_R$ (0) for the remote
(background) impurities. We assume the random quenched charged
impurities to all have unit strength (i.e., having an elementary
charge of $\pm e$ each) with no loss of generality.

Our model thus has four independent parameters with dimensions of
length: $n^{-1/2}$, $n_R^{-1/2}$, $n_B^{-1/2}$, $d_R$. In addition to
these (experimentally variable) parameters, we also have $m$ and
$\kappa$ defining the material system which is fixed in all samples
for a given material. In principle, two additional materials
parameters should be added to describe the most general situation,
namely, the spin ($g_s$) and the valley ($g_v$) degeneracy, but we
assume $g_s=2$ and $g_v=1$ throughout the current work (and for all
our numerical results) since our interest here is entirely focussed on
high-mobility n- and p-GaAs 2D systems 
where the mobility/quality dichotomy has mostly been
discussed. Additional experimentally relevant (but, theoretically
non-essential) parameters, such as a finite width of the 2D electron
system (instead of the strict 2D limit) and/or a 3D distribution of
the random impurities (instead of the 2D distribution assumed above),
are straightforward to include in the model and are not discussed
further in details.

The mobility (quality) is now simply defined by the characteristic
scattering time $\tau_t$ ($\tau_q$) as given by the following
equations in our leading-order transport theory (i.e., Boltzmann
transport plus Born approximation for scattering):  
\begin{eqnarray}
\frac{1}{\tau_t(k)} = \frac{2\pi}{\hbar} \sum_{\bf k'} \sum_l &
&  \int_{-\infty}^{\infty} dz N_i^{(l)}(z)  \left |u_{\bf k-k'}(z) \right
  |^2 \nonumber  \\ 
& &\times   (1-\cos\theta) \delta(E_{\bf k}-E_{\bf k'}),
\label{eq1}
\end{eqnarray}
and
\begin{eqnarray}
\frac{1}{\tau_q(k)} = \frac{2\pi}{\hbar} \sum_{\bf k'} \sum_l &
&  \int_{-\infty}^{\infty} dz N_i^{(l)}(z)  \left |u_{\bf k-k'}(z) \right
  |^2 \nonumber  \\ 
& &\times   \delta(E_{\bf k}-E_{\bf k'}),
\label{eq2}
\end{eqnarray}
where $N_i^{(l)}(z)$ is the 3D impurity distribution for the $l$-th
kind of disorder in the system, and $u_{\bf q}(z)$ is the screened
electron-impurity Coulomb interaction given by:
\begin{equation}
u_{\bf q}(z) = \frac{V_{\bf q}(z)}{\varepsilon(q)} = \frac{2\pi e^2}{\kappa q}
\frac{e^{-qz}}{\varepsilon(q)},
\label{eq3}
\end{equation}
with $\varepsilon(q)$ being the static RPA dielectric function for the
2D electron system. We note that $V_q(z) = V(q) e^{-qz} = \frac{2\pi
  e^2}{\kappa q} e^{-qz}$ is simply the 2D Fourier transform of the 3D
$1/r$ Coulomb potential, which explicitly takes into account the fact
that a spatial separation of `$z$' may exist between the 2D electron
layer and the charged impurities. The 2D static RPA dielectric
function or the screening function is given by
\begin{equation}
\varepsilon(q) = 1 + \frac{2\pi e^2}{\kappa q} \Pi(q),
\label{eq4}
\end{equation}
where the static 2D electronic polarizability function $\Pi(q)$ is
given by
\begin{equation}
\Pi(q) = N_F \left [ 1- \theta(q-2k_F) \sqrt{1-(2k_F/q)^2} \right ],
\label{eq5}
\end{equation}
where $N_F = m/\pi \hbar^2$ and $\theta(x) = 0$ (1) for $x <0$ ($x>0$)
is the Heaviside step (or theta) function and the 2D Fermi wave vector
$k_F$ is determined by the 2D carrier density through the formula $k_F
= (2\pi n)^{1/2}$. We note that the 2D Fermi energy ($E_F$) is given
by $E_F = \hbar^2k_F^2/2m = \pi \hbar^2 n/m$. The RPA screening
function $\varepsilon(q)$ can be expressed in the convenient form
\begin{equation}
\varepsilon(q) = 1 + q_s/q,
\end{equation}
which is exactly equivalent to Eqs.~(\ref{eq4}) and (\ref{eq5}) if we
define the 2D screening wave vector $q_s$ to be
\begin{equation}
q_s = q_{TF} \left [ 1- \theta(q-2k_F) \sqrt{1-(2k_F/q)^2} \right ],
\label{eq8}
\end{equation}
where the 2D Thomas-Fermi wave vector $q_{TF}$ is defined to be
\begin{equation}
q_{TF} = 2 m e^2/(\kappa \hbar^2).
\end{equation}
We note that (1) the Thomas-Fermi wave vector $q_{TF}$ is proportional
to the 2D density of states at the Fermi energy $N_F = m/\pi \hbar^2$
(and is inversely proportional to the effective background lattice
dielectric constant $\kappa$), and (2) screening is constant in 2D for
$0 \leq q \leq 2k_F$ [see Eq.~(\ref{eq8})] because of the constant
energy-independent 2D electronic density of states. Since the
$\delta$-functions (necessary for energy conservation during the
impurity-induced elastic scattering of an electron from the momentum
state $|{\bf k}\rangle$ to the momentum state $|{\bf k}' \rangle$ with
a net wave vector transfer of ${\bf q} = {\bf k} - {\bf k}'$) in
Eqs.~(\ref{eq1}) and (\ref{eq2}) restrict the wave vector transfer $0
\leq q \leq 2k_F$ range (this is simply because we are at $T=0$ so
that the maximum possible scattering wave vector is $2k_F$
corresponding to the pure back-scattering of an electron from $+k_F$
to $-k_F$ while obeying energy conservation), the relevant screening
wave vector for our problem is purely the Thomas-Fermi wave vector
$q_s = q_{TF}$ as follows from Eq.~(\ref{eq8}) for $q\le 2k_F$.

We can, therefore, rewrite Eq.~(\ref{eq3}) as
\begin{equation}
u_q(z) = \frac{2\pi e^2}{\kappa (q+ q_{TF})} e^{-qz},
\label{eq10}
\end{equation}
giving the effective screened Coulomb interaction (in the 2D momentum
space) between an electron in the 2D layer and a random quenched
charged impurity located a distance `$z$' away. 

Finally, our 2-impurity disorder model is given by
\begin{equation}
N_i(z) = n_R \delta(z-d_R) + n_B \delta(z),
\label{eq11}
\end{equation}
with three independent parameters $n_R$, $d_R$, and $n_B$ completely
defining the underlying disorder. We note that writing the second term
in Eq.~(\ref{eq11}) as $n_B \delta(z-d_B)$, and thus introducing an
additional length parameter into the model, is completely unnecessary
since, as we will see below, the physics of the mobility versus
quality duality in high-quality 2D structures is entirely dominated by
scattering from near impurities (controlling mobility) and far
impurities (controlling quality), which allows us to put $d_B=0$ in
the minimal model (thus making the near impurities very near
indeed). Putting Eqs.~(\ref{eq3}) -- (\ref{eq11}) in Eqs.~(\ref{eq1})
and (\ref{eq2}), we can combine them into a single 2D integral,
obtaining 
\begin{eqnarray}
\frac{1}{\tau_{t,q}} = \frac{2\pi}{\hbar} \left (\frac{2\pi
  e^2}{\kappa} \right )^2 & &  
\int\frac{d^2k'}{(2\pi)^2} \frac{\delta(E_{\bf k}-E_{\bf k'})}{(q_{TF}
  + |{\bf k}-{\bf k'}| )^2} 
 f_{t,q}(\theta) \nonumber  \\ 
& &\times   \left \{ n_R e^{-2|{\bf k}-{\bf k'}| d_R} + n_B \right \},
\label{eq12}
\end{eqnarray}
where $f_t(\theta) = 1- \cos \theta_{kk'}$ and $f_q(\theta) =1$. For
completeness, we mention that $E_{\bf k} = \hbar^2 k^2/2m$.

Before proceeding further, we emphasize that the ``only" difference
between mobility (i.e., $\tau_t$) and quality (i.e., $\tau_q$) is the
appearance (or not) of the angular factor $(1-\cos \theta)$ inside the
double integral in Eq.~(\ref{eq12}) for $\tau_t$ ($\tau_q$). This
arises from the fact that the mobilty or the conductivity is
unaffected by forward scattering (i.e., $\theta \approx 0$ or $\cos
\theta \approx 1$) whereas the single-particle level-broadening
($\Gamma \sim
\hbar/\tau_q$) is sensitive to scattering through all
angles. Technically, the ($1-\cos\theta$) factor arises from the
impurity scattering induced vertex correction in the 2-particle
current-current correlation function representing the electrical
conductivity whereas the single-particle scattering rate
($\tau_q^{-1}$) is given essentially by the imaginary part of the
impurity scattering induced 1-particle electronic self-energy which
does not have any vertex correction in the leading-order impurity
scattering strength. The absence (presence) of the vertex correction
in $\tau_q^{-1}$ ($\tau_t^{-1}$) makes the relevant scattering rate
sensitive (insensitive) to forward scattering, leading to a situation
where $\tau_q$ and $\tau_t$ could be very different from each other if
forward (or small angle) scattering is particularly  important in a
system as it would be for long range disorder potential.

This could happen in 3D systems if the scattering potential is
strongly spatially asymmetric (or non-spherical) for some reason. It
was pointed out \cite{dassarmaPRB1985} a long time ago and later
experimentally verified \cite{exp11} that such a strongly
non-spherically symmetric scattering potential arises naturally in 2D
modulation-doped structures from random charged impurities placed very
far ($k_Fd \gg 1$) away from the 2D electron system due to the
influence of the exponential $e^{-q d}$ factor in the Coulomb
potential which restricts much of the scattering to $q \ll 1/z$, thus
exponentially enhancing the importance of small-angle (i.e., small
scattering wave vector) scattering. This then leads to $\tau_t \gg
\tau_q$ for scattering dominated by remote dopants (the two scattering
times could differ by more than two orders of magnitude in
high-mobility modulation-doped structures where $k_F d_R \gg 1$ is
typically satisfied due to the far-away placement of the remote
dopants in order to minimize large-angle resistive scattering
processes) in 2D systems, but for the unintentional background
impurities, which always satisfy $k_F d_B \ll 1$ by definition (since
$d_B \approx 0$), the two scattering times are approximately equal.  
We note that the finite layer thickness of the 2D system puts a lower
bound on how small $d_B$ can be, but this has no qualitative
significance for our consideration.

Now, we immediately realize the crucial relevance of the 2-impurity
model in distinguishing mobility (i.e., $\tau_t$) and quality (i.e.,
$\tau_q$) in 2D systems since it now becomes possible for one type of
disorder (e.g., remote impurities) to control the quality (i.e.,
$\tau_q$) and the other type to control the mobility (i.e.,
$\tau_t$). Of course, whether such a distinction actually applies to a
given situation or not depends entirely on the details of the sample
parameter (i.e., the specific values of $n_R$, $d_R$, $n_B$) as well
as the carrier density $n$, but the possibility certainly exists for
$d_R$ to be large enough so that the remote impurity scattering is
almost entirely small-angle scattering (thus only adversely affecting
$\tau_q$ in an appreciable way) whereas the background impurity
scattering determines $\tau_t$. If this happens, then mobility
($\tau_t$) and quality ($\tau_q$) could very well be very different in
2D samples, and may have little to do with each other.

To bring out the above physical picture explicitly, we now provide
some analytical calculations for the integrals in Eq.~(\ref{eq12})
defining mobility ($\tau_t$) and quality ($\tau_q$).  We rewrite
Eq.~(\ref{eq12}) as 
\begin{equation}
\tau^{-1}_{t,q} = I^{(R)}_{t,q} + I^{(B)}_{t,q},
\label{eq16}
\end{equation}
where
\begin{eqnarray}
I^{(R)}_{t,q}  = n_RV_0  \frac{2\pi}{\hbar} & & 
\int\frac{d^2k'}{(2\pi)^2} \frac{\delta(E_{\bf k}-E_{\bf k'})}{(q_{TF} + |{\bf k}-{\bf k'}| )^2}
 \nonumber  \\ 
& &\times   f_{t,q}(\theta)  e^{-2|{\bf k}-{\bf k'}| d_R},
\label{eq17}
\end{eqnarray}
and
\begin{equation}
I^{(B)}_{t,q}  = n_B V_0  \frac{2\pi}{\hbar}
\int\frac{d^2k'}{(2\pi)^2} \frac{\delta(E_{\bf k}-E_{\bf k'})}{(q_{TF} + |{\bf k}-{\bf k'}| )^2}
 f_{t,q}(\theta),
\label{eq18}
\end{equation}
where $V_0=(2\pi e^2/\kappa)^2$.
Eqs.~(\ref{eq17}) and (\ref{eq18}) can be rewritten in dimensionless forms as
\begin{equation}
I^{(R)}_{t,q} = \left ( \frac{mn_R V_0}{2\pi \hbar^3 k_F^2} \right )  \int_0^1 dx \frac{g_{t,q}(x) e^{-2 x a_R}}{\sqrt{1-x^2}(x+s)^2}
\label{eq19}
\end{equation}
and
\begin{equation}
I^{(B)}_{t,q}  = \left ( \frac{mn_B V_0}{2\pi \hbar^3 k_F^2} \right ) 
 \int_0^1 dx \frac{g_{t,q}(x)}{\sqrt{1-x^2}(x+s)^2}
\label{eq20}
\end{equation}
where $a_R \equiv 2k_F d_R$ and $s=q_{TF}/2k_F$, and $g_t(x) = 2x^2$, $g_q(x) = 1$.

To proceed further analytically, we now make the assumption that, by definition, the remote dopants are far enough that the condition $a_R = 2k_F d_R \gg 1$ is satisfied. We note that for any arbitrarily large value of $d_R$, this ``remote" impurity condition breaks down at a low enough carrier density such that $n \alt 1/ 8\pi d_R^2$. Thus, the distinction between `far' and `near' impurities in our 2-impurity model starts disappearing at very low carrier density. Putting $a_R \gg 1$ as well as $s=q_{TF}/2k_F \gg 1$ we can obtain the following asymptotic expressions for $I_{t,q}^{(R)}$
\begin{equation}
I^{(R)}_{t} = \left ( \frac{m n_R}{2\pi \hbar^3 k_F^2} \right ) \left ( \frac{2\pi e^2}{\kappa} \right )^2
\left ( \frac{2}{3s^2} \right ) \frac{1}{a_R^3},
\label{eq24}
\end{equation}
\begin{equation}
I^{(R)}_{q} = \left ( \frac{m n_R}{2\pi \hbar^3 k_F^2} \right ) \left ( \frac{2\pi e^2}{\kappa} \right )^2
\frac{1}{a_R s^2},
\label{eq25}
\end{equation}
\begin{equation}
I^{(B)}_{t} = \left ( \frac{m n_B}{2\pi \hbar^3 k_F^2} \right ) \left ( \frac{2\pi e^2}{\kappa} \right )^2
\frac{2\pi}{s^2},
\label{eq26}
\end{equation}
\begin{equation}
I^{(B)}_{q} = \left ( \frac{m n_B}{2\pi \hbar^3 k_F^2} \right ) \left ( \frac{2\pi e^2}{\kappa} \right )^2
\frac{2\pi}{s^2},
\label{eq27}
\end{equation}
with $\tau_{t,q}^{-1} = I_{t,q}^{(R)} + I_{t,q}^{(B)}$.
Equations (\ref{eq24}) -- (\ref{eq27}) provide us with approximate analytical expressions for the contributions $I_{t,q}^{(R)}$ and $I_{t,q}^{(B)}$ by the remote and the background
scattering respectively to the transport and quantum scattering rates. We note that in these asymptotic limits ($a_R \gg 1$ and $s \gg 1$)
\begin{equation}
\frac{I_t^{(B)}}{I_t^{(R)}} = 3 \pi a_R^3 \frac{n_B}{n_R},
\label{eq29}
\end{equation}
and
\begin{equation}
\frac{I_q^{(R)}}{I_q^{(B)}} = \frac{1}{2\pi a_R}\frac{n_R}{n_B}.
\label{eq30}
\end{equation}
In addition,
\begin{equation}
{I_q^{(R)}}/{I_t^{(R)}} = 3 a_R^2/2,
\label{eq31}
\end{equation}
\begin{equation}
{I_q^{(B)}}/{I_t^{(B)}} = 1.
\label{eq32}
\end{equation}
We note, therefore, that $I_q^{(R)} \gg I_t^{(R)}$ (since $a_R \gg 1$)
as is already well established, and that $I_t^{(B)} > I_t^{(R)}$ if
$3\pi a_R^3 > n_R/n_B$, whereas $I_q^{(R)} > I_q^{(B)}$ if $n_R/n_B >
2 \pi a_R$. Consistency demands that $3\pi a_R^3 \gg 2 \pi a_R$, i.e.,
$a_R^2 \gg 2/3$, which is guaranteed since $a_R \gg 1$.

It is thus possible for the $B$-scatterers to dominate $\tau_t^{-1}$,
i.e., $I_t^{(B)} \gg I_t^{(R)}$, and $R$-scatterers to dominate
$\tau_q^{-1}$, i.e., $I_q^{(R)} \gg I_q^{(B)}$ if the following
conditions are both satisfied 
\begin{eqnarray}
3 \pi a_R^3 & \gg & n_R/n_B, \nonumber \\
2 \pi a_R & \ll & n_R/n_B.
\label{eq35}
\end{eqnarray}
The two inequalities defined by Eq.~(\ref{eq35}) are not mutually exclusive if $n_R \gg n_B$ and 
\begin{equation}
n_R/2\pi n_B \gg a_R \gg (n_R/3\pi n_B)^{1/3}.
\label{eq36}
\end{equation}
It is easy to see the Eq.~(\ref{eq36}) is consistent as long as $n_R
\gg 5.13 n_B$, a perfectly reasonable scenario! 
In fact, we expect this condition to be extremely well-satisfied in
high-quality 2D systems where $n_B$ is very small, but $n_R \approx n$
due to modulation doping and charge neutrality.

The above analytic considerations lead to the conclusion that it is
possible for $\tau_t^{-1}$ to be dominated by background impurities,
and at the same time for $\tau_q^{-1}$ to be dominated by the remote
impurities provided the necessary conditions $a=2k_Fd_R \gg 1$ and
$n_R \gg 5.13 n_B$ are obtained. We emphasize that these are only {\it
  necessary} 
conditions, and not sufficient conditions. Whether the 2-impurity
model indeed leads to mobility (i.e., $\tau_t$) and quality (i.e.,
$\tau_q$) being controlled by physically distinct disorder mechanisms
in real 2D semiconductor structures can only be definitely established
through explicit numerical calculations of $\tau_{t,q}^{-1}$ for
specific disorder configurations, which we do in the next section of
this article. The general analytical theory developed above also
explicitly shows that there can be no mobility/quality dichotomy  if
there is only one type of disorder mechanism operational in the sample
since both quality and mobility will then be controlled by exactly the
same disorder parameters (unlike the situation discussed above).

It may be useful to write a full analytical expression for $\tau_t^{-1}$ and $\tau_q^{-1}$ (in the $a_R \gg 1$ limit) combining all the expressions given above to show how the disorder parameters $n_R$, $d_R$, and $n_B$ enter the expressions for mobility and quality:
\begin{equation}
\tau_t^{-1} = I_t^{(R)} + I_t^{(B)} = A_t^{(R)}n_R/a_R^3 + A_t^{(B)} n_B,
\label{eq37}
\end{equation}
\begin{equation}
\tau_q^{-1} = I_q^{(R)} + I_q^{(B)} = A_q^{(R)}n_R/a_R + A_q^{(B)} n_B,
\label{eq38}
\end{equation}
where
\begin{eqnarray}
A_t^{(R)} & = & \left ( \frac{m}{2\pi\hbar^3} \right ) \left ( \frac{2\pi e^2}{\kappa} \right )^2 \left ( \frac{8}{3 q_{TF}^2} \right ) \nonumber \\
A_t^{(B)} & = & \left ( \frac{m}{2\pi \hbar^3} \right ) \left ( \frac{2\pi e^2}{\kappa} \right )^2 \left ( \frac{8\pi}{q_{TF}^2} \right ),
\label{eq39}
\end{eqnarray}
\begin{eqnarray}
A_q^{(R)} & = & \left ( \frac{m}{2\pi\hbar^3} \right ) \left ( \frac{2\pi e^2}{\kappa} \right )^2 \left ( \frac{4}{q_{TF}^2} \right ) \nonumber \\
A_q^{(B)} & = & \left ( \frac{m}{2\pi \hbar^3} \right ) \left ( \frac{2\pi e^2}{\kappa} \right )^2 \left ( \frac{8\pi}{q_{TF}^2} \right ).
\label{eq40}
\end{eqnarray}
Equations (\ref{eq37}) and (\ref{eq38}) immediately lead to the approximate sufficient conditions for mobility and quality to be controlled by background and remote impurities, respectively:
\begin{equation}
n_B \gg n_R/a_R^3, \;\; i.e., \; n_R \ll n_B a_R^3,
\label{eq41}
\end{equation}
and
\begin{equation}
n_R/a_R \gg n_B, \;\; i.e., \; n_R \gg n_B a_R.
\label{eq42}
\end{equation}
Since $a_R \gg 1$ by definition, the two conditions, Eqs.~(\ref{eq41})
and (\ref{eq42}), can simultaneously be satisfied if
\begin{equation}
a_R^3 \gg n_R/n_B \gg a_R,
\label{eq43}
\end{equation}
with $a_R \gg 1$. Equation (\ref{eq43}) gives the sufficient condition
for the existence of a mobility/quality dichotomy in 2D semiconductor
structures. Since the unintentional background charged impurity
concentration is typically very low in high-mobility 2D GaAs
structures and since the remote charged dopant density is typically
(at least) equal to the carrier density, the condition $a_R^3 \gg
n_R/n_B \gg a_R \gg 1$ can certainly be satisfied in some 2D samples
(but obviously not in all samples). For example, a typical
modulation-doped high mobility 2D GaAs-Al$_x$Ga$_{1-x}$As structure
may have $n \approx n_R \approx 3 \times 10^{11}$ cm$^{-2}$; $d_R
\approx 1000$ \AA; $n_B \approx 10^8$ cm$^{-2}$ (corresponding roughly
to a 3D bulk background charged impurity density of $3\times 10^{13}$
cm$^{-3}$ for a 300 \AA \; wide quantum well structure). These system
parameters satisfy the constraint defined by Eq.~(\ref{eq43}) with
$k_Fd_R \approx 15$, i.e., $a_R \approx 30$; $n_R/n_B \approx
3000$. We emphasize the obvious role of carrier density here, i.e.,
lowering the carrier density decreases $k_F d$ (and hence $a_R$), and
eventually at low enough carrier density, $\tau_t^{-1}$ and
$\tau_q^{-1}$ are determined by the same disorder parameters!

In concluding this theoretical section, let us consider a concrete numerical example of the mobility/quality dichotomy using two hypothetical samples (1 and 2) with the following realistic sample parameters:
\begin{eqnarray}
{\rm Sample \; 1:}& &  \; d_R^{(1)} = 500 \AA; \; n_B^{(1)} = 10^7 cm^{-2} \nonumber \\
{\rm Sample \;2:}& &  \; d_R^{(1)} = 1000 \AA; \; n_B^{(2)} = 10^8 cm^{-2}. 
\label{eq44}
\end{eqnarray}  
We assume the sample carrier density to be the same in both cases (so that ``an apple-to-apple" comparison in being made): $n=4\times 10^{11}$ cm$^{-2}$. For the purpose of keeping the number of parameters a minimum we assume $n_R = n = 4\times 10^{11}$ cm$^{-2}$ also for both samples. Using the analytical theory developed above (or by direct numerical calculations), we find
\begin{equation}
\tau_t^{(1)}/\tau_t^{(2)} \approx 5; \; \tau_q^{(1)}/\tau_q^{(2)} \approx 2.
\label{eq45}
\end{equation}
This means that sample 1 (with $n_B^{(1)} < n_B^{(2)}$) has a five
times higher mobility than sample 2 whereas sample 2 (with $d_R^{(2)}
> d_R^{(1)}$) has two times higher `quality' than sample 1, i.e.,
sample 2 has a single-particle level broadening $\Gamma$ (or ``Dingle
temperature") which is half of that of sample 1 although both samples
have exactly the same carrier density! This realistic example shows
that it is generically possible in the 2-impurity model for $\mu_1 >
\mu_2$ and $\Gamma_1 > \Gamma_2$, with the conclusion that a higher
mobility does not necessarily ensure a higher quality. We emphasize
that (1) this would not be possible within the 1-impurity model, and
(2) this conclusion is density dependent -- for much lower carrier
density, where $k_F d \gg 1$ condition cannot be satisfied for the
remote dopants, mobility and quality will again be closely connected
since at sufficiently low carrier density, the 2-impurity model
effectively reduces to an 1-impurity model.

\begin{figure}[t]
	\centering
	\includegraphics[width=1.\columnwidth]{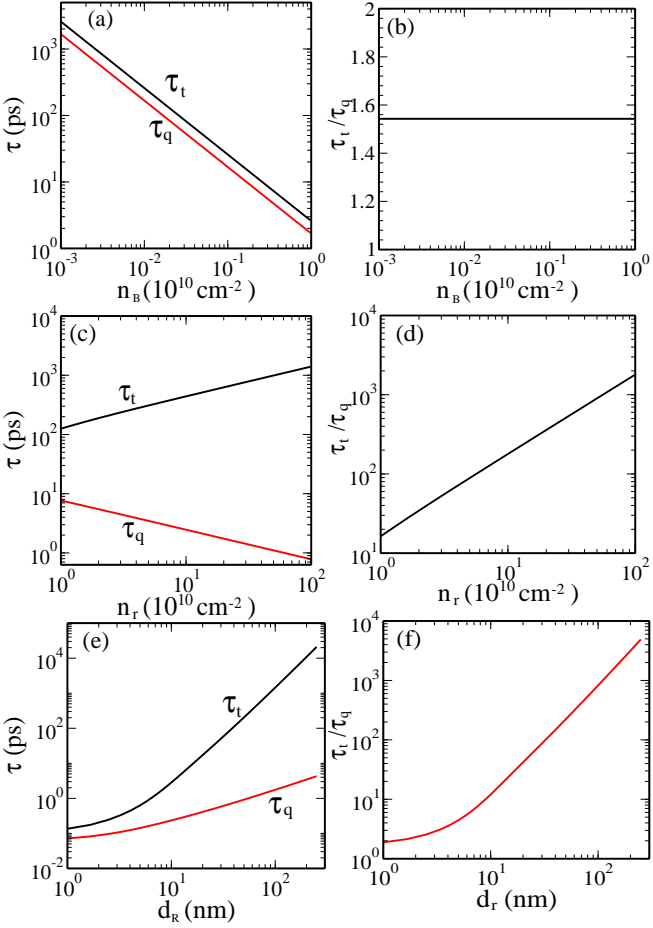}
	\caption{(a) The scattering times ($\tau_t$ and $\tau_q$) and (b) the ratio ($\tau_t/\tau_q$) as a function of the background impurity density $n_B$ for $n=3\times 10^{11} cm^{-2}$ and $n_R=0$. 
(c) The scattering times and (d) the ratio ($\tau_t/\tau_q$) as a function of the remote impurity density $n_R$ for $n=n_R$, $n_B= 0$, and $d_R=80$nm.
(e) The scattering times and (f) the ratio ($\tau_t/\tau_q$)  as a function of the impurity location $d_r$ for $n=n_r=3\times 10^{11} cm^{-2}$ and the background impurity density $n_B=0$.
Here the $\delta$-layer is considered (i.e., $a=0$).  
}
\label{fig01}
\end{figure}

\begin{figure}[t]
	\centering
	\includegraphics[width=1.\columnwidth]{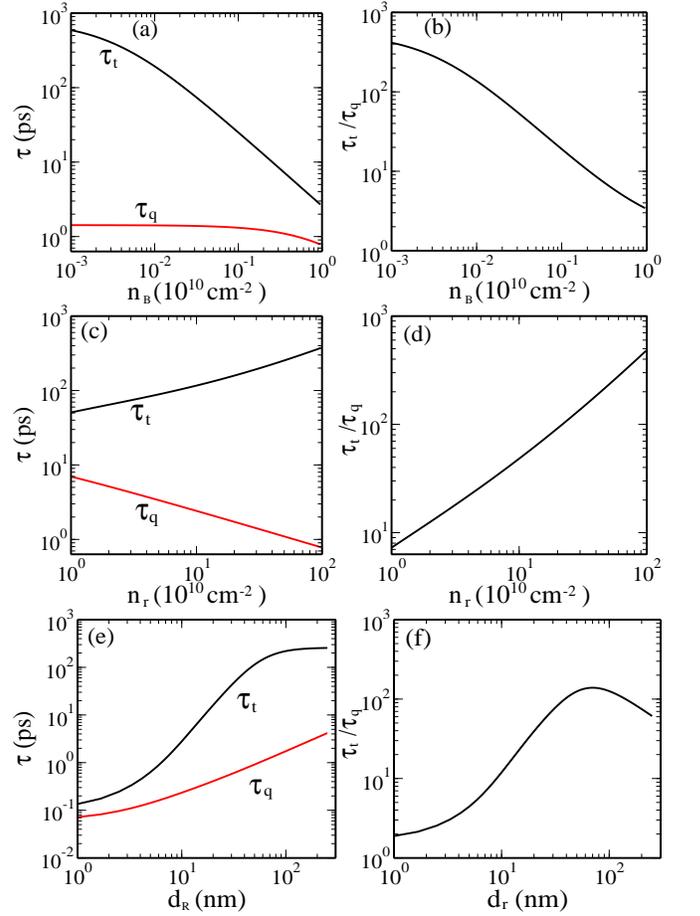}
	\caption{(a) The scattering times ($\tau_t$ and $\tau_q$) and (b) the ratio ($\tau_t/\tau_q$) as a function of the background impurity density $n_B$ for $n=n_R=3\times 10^{11} cm^{-2}$ and $d_R=80$ nm. 
(c) The scattering times and (d) the ratio ($\tau_t/\tau_q$) as a function of the remote impurity density $n_R$ for $n=n_R$, $n_B= 10^8$ cm$^{-2}$, and $d_R=80$nm.
(e) The scattering times and (f) the ratio ($\tau_t/\tau_q$)  as a function of the impurity location $d_R$ for $n=n_R=3\times 10^{11} cm^{-2}$ and the background impurity density $n_B=0$.
Here the $\delta$-layer is considered (i.e., $a=0$).  
}
\label{fig02}
\end{figure}

\begin{figure}[t]
	\centering
	\includegraphics[width=1.\columnwidth]{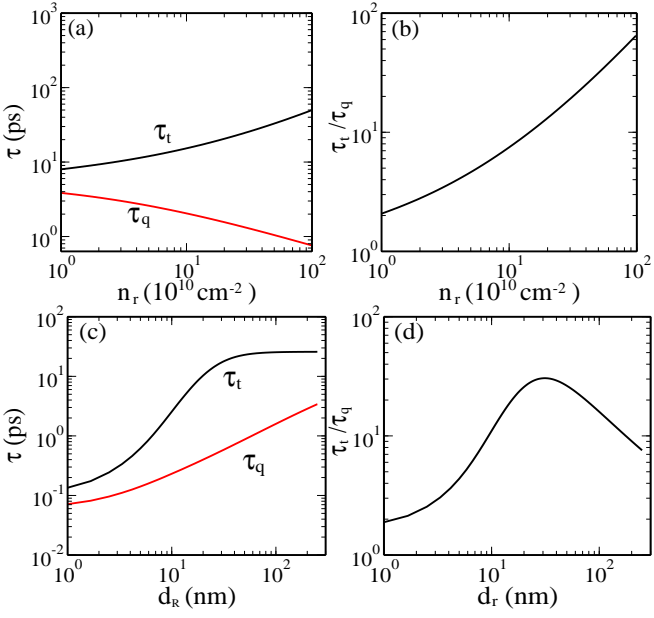}
	\caption{(a) The scattering times and (b) the ratio ($\tau_t/\tau_q$) as a function of the remote impurity density $n_R$ for $n=n_R$, $n_B= 10^9$ cm$^{-2}$, and $d_R=80$nm.
(c) The scattering times and (d) the ratio ($\tau_t/\tau_q$)  as a function of the impurity location $d_R$ for $n=n_R=3\times 10^{11} cm^{-2}$ and the background impurity density $n_B=10^9$ cm$^{-2}$.
Here the $\delta$-layer is considered (i.e., $a=0$).  
}
\label{fig03}
\end{figure}

\begin{figure}[t]
	\centering
	\includegraphics[width=1.\columnwidth]{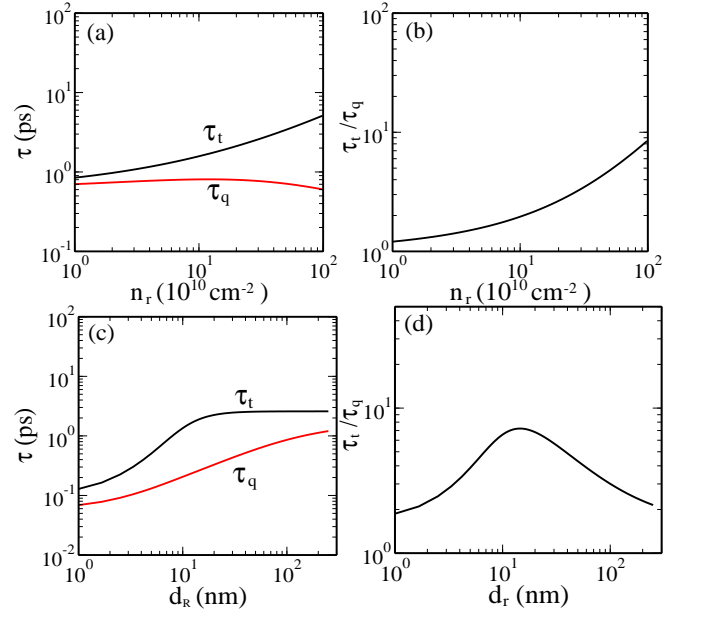}
	\caption{
(a) The scattering times and (b) the ratio ($\tau_t/\tau_q$) as a function of the remote impurity density $n_R$ for $n=n_r$, $n_B= 10^{10}$ cm$^{-2}$, and $d_R=80$nm.
(c) The scattering times and (d) the ratio ($\tau_t/\tau_q$)  as a function of the impurity location $d_R$ for $n=n_r=3\times 10^{11} cm^{-2}$ and the background impurity density $n_B=10^{10}$ cm$^{-2}$.
Here the $\delta$-layer is considered (i.e., $a=0$).  
}
\label{fig04}
\end{figure}

Before presenting our realistic numerical results for 2D GaAs-AlGaAs
structures (including both the quasi-2D finite well-width effect and a
3D distribution of the background unintentional charged impurity
distribution within the well) using the `near' and `far' 2-impurity
model in the next section (Sec. III), we conclude the current theory
section by showing some numerical results for $\tau_t$ and $\tau_q$
using the idealized 2-impurity model [i.e., Eq.~(\ref{eq11})] and the
strict 2D model for the semiconductor structure. These results
presented in Figs.~\ref{fig01} -- \ref{fig04} explicitly visually
demonstrate that $\tau_t$ and $\tau_q$ cannot be single-valued
functions of each other as long as the underlying disorder consists of
(at least) two distinct scattering mechanisms as operational within the
2-impurity model. The results shown in Figs.~\ref{fig01} --
\ref{fig04} also serve to establish the validity of the analytical
theory we provided above in this section.

In Fig.~\ref{fig01} we show the dependence of the calculated $\tau_t$
and $\tau_q$ on the individual scattering mechanism (i.e., the
near-impurity scattering strength defined by the background impurity
concentration $n_B$ or the far-impurity scattering strength defined by
either $n_R$ or $d_R$) assuming that the other mechanism is absent
(i.e., just an effective 1-impurity model applies). Results in
Fig.~\ref{fig01} should be compared with the corresponding results in
Figs.~\ref{fig02} -- \ref{fig04} where both scattering mechanisms are
operational to clearly see that $\tau_t$ and $\tau_q$ are manifestly
not unique functions of each other by any means and a given value of
$\tau_t$ (or $\tau_q$) could lead to distinct values of $\tau_q$ (or
$\tau_t$) depending on the details of the disorder distribution. Thus,
mobility ($\tau_t$) and quality ($\tau_q$) are not simply connected.

In presenting the results for Figs.~\ref{fig01} -- \ref{fig04} we
first note that $\tau_{t,q} = \tau(n,n_B,n_R,d_R)$ even within the
simple 2-impurity model. Since the carrier density dependence of
$\tau$ is not the central subject matter of our interest in the
current work (and has been discussed elsewhere by us
\cite{dassarmaprb2013}, we simplify the presentation by assuming
$n_R=n$ in Figs.~\ref{fig01} -- \ref{fig04} which also assures a
straightforward charge neutrality. This, however, has important
implications since the dependence on $n_R$ and $n$ now become
compounded rather than being independent. Thus the $n_R$-dependence of
$\tau$ shown in Figs.~\ref{fig01} -- \ref{fig04} is not the trivial
$\tau_{t,q} \sim n_R^{-1}$ behavior (as it is for the
$n_B$-dependence, where $\tau_{t,q} \sim n_B^{-1}$) since the $n_R =
n$ condition synergistically combines both $n_R$ and $n$
dependence. We mention that in ungated samples with fixed carrier
density, the condition $n=n_R$ is perfectly reasonable, and therefore,
the results shown in Figs.~\ref{fig01} -- \ref{fig04} apply to
modulation-doped samples with fixed carrier density $n=n_R$.

From Fig.~\ref{fig01}, we immediately conclude the obvious: The
functional relationship between $\tau_t$ and $\tau_q$ depends entirely
on which disorder parameter is being varied -- in fact
Figs.~\ref{fig01}(a), (c), (e) give three completely distinct
functional relationships between $\tau_t$ and $\tau_q$ depending on
whether $n_B$, $n_R$, or $d_R$ is being varied, respectively. From the
corresponding values of $\tau_t/\tau_q$, as shown in
Figs. \ref{fig01}(b), (d), (f) respectively we clearly see that
$\tau_t$ and $\tau_q$ variations with individual disorder parameters
$n_B$, $n_R$, $d_R$ are very different indeed. We point out an
important qualitative aspect of Fig.~\ref{fig01}(c) which has not been
explicitly discussed in the literature and which has important
implications for the mobility/quality dichotomy. Here the mobility
(i.e., $\tau_t$) increases with increasing $n_R = n$, but the quality
(i.e., $\tau_q$) decreases with increasing $n=n_R$. This is a peculiar
feature of long range Coulomb scattering by remote dopants.

Figs.~\ref{fig02} -- \ref{fig04} explicitly show how the 2-impurity
model can very strongly modify the 1-impurity model functional
dependence of $\tau_{t,q}$ on the disorder parameters $n_B$, $n_R$
($=n$), and $d_R$. Clearly, depending on the specific 2D samples,
$\tau_t$ and $\tau_q$ could behave very differently as already
established in our analytical theoretical results given above. For
example, in contrast to Fig.~\ref{fig01}(a), where both $\tau_t$ and
$\tau_q$ decrease monotonically (and trivially as $n_B^{-1}$) with
increasing amount of unintentional background impurity density $n_B$,
Fig.~\ref{fig02}(a) shows that
$\tau_q$ is essentially a constant whereas $\tau_t$ decreases with
increasing $n_B$, thus demonstrating a specific example of how the
effective quality (i.e. $\tau_q$) remains the same although the
effective mobility decreases by more than an order of magnitude due to
the variation in the background disorder. Similarly, in
Fig.~\ref{fig01}(c), increasing the remote dopant separation $d_R$
(keeping $n=n_R$ fixed, and $n_B=0$) increases both $\tau_t$ and
$\tau_q$ monotonically (with $\tau_t$ increasing as $d^3$ in contrast
to $\tau_q$ increasing as $d$ for large $d$), but Fig.~\ref{fig02}(e),
\ref{fig03}(c), \ref{fig04}(c) show that, depending on the background
impurity scattering strength, $\tau_t$ basically saturates for larger
$d$ since it is then dominated by the unintentional background
impurities rather than by the remote dopants whereas $\tau_q$
continues to be limited by the remote dopants. This leads to an
interesting nonmonotonicity in $\tau_t/\tau_q$ as a function of $d_R$
in Figs.~\ref{fig02} -- \ref{fig04} in contrast to Fig.~\ref{fig01}(c)
where $\tau_t/\tau_q \sim d_R^2$ keeps on increasing forever in the
absence of background scattering. The realistic dependence of
$\tau_{t,q}$ on $n_R$ (with $n=n_R$) in the presence of fixed $n_B$
and $d_R$ remains qualitatively similar in Figs.~\ref{fig01} --
\ref{fig04} although there could be large quantitative differences,
indicating that increasing $n_R$ ($=n$) would typically by itself tend
to enhance (suppress) $\tau_t$ ($\tau_q$), but the effect becomes
whether for larger (smaller) values of $n_B$ ($d_R$). This is an
important result of our paper.

The analytical and numerical results presented in this section
establish clearly that $\tau_t$ and $\tau_q$ can essentially be
independent functions of the disorder parameters in the 2-impurity
model, and thus, mobility and quality could, in principle, have little
to do with each other in realistic 2D semiconductor structures. We
make this point even more explicit by carrying out calculations in
experimentally realistic samples in the next section of this article.

\section{Numerical results and discussions}

\begin{figure}[t]
	\centering
	\includegraphics[width=1.\columnwidth]{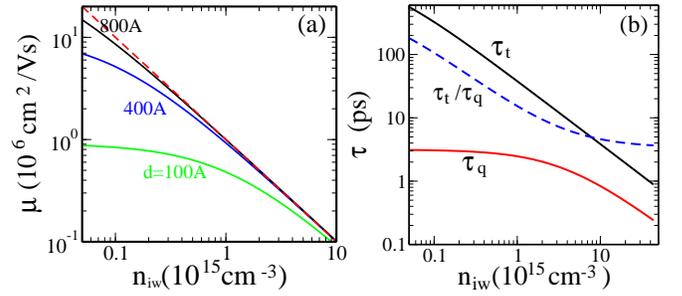}
	\caption{(a)
Calculated mobilities as a function of background 3D charged impurity density $n_{iw}$ for
fixed remote impurity density, $n_i=1.5\times 10^{11}cm^{-2}$, and
electron density, $n=3\times 10^{11}cm^{-2}$. The red dashed line
indicates $\mu \sim n_{iw}$.
The numbers indicate the location of 
remote impurities  which is measured from the interface of quantum well. 
(b) The calculated scattering times and their ratio
$\tau_t/\tau_q$ as a function of background
impurity density using the same parameters of (a) and $d=800$ \AA.
The dashed line represents the ratio of $\tau_t$ to $\tau_q$. 
A quantum well with the thickness of $a=300$ \AA \;  is used in this calculation.  
	}
	\label{fig1}
\end{figure}

We begin presenting our realistic numerical results for 2D transport
properties (both $\tau_t$ or mobility and $\tau_q$ or quality) without
any reference to the analytical asymptotic theoretical results presented in the
last section by showing in Fig.~\ref{fig1} the calculated $T=0$
mobility ($\mu$), transport scattering time ($\tau_t$) and the
single-particle (or quantum) scattering time ($\tau_q$) for a 2D
GaAs-AlGaAs sample at a fixed carrier density ($n=3\times 10^{11}$
cm$^{-2}$) in the presence of two types of disorder: a remote charged
impurity sheet ($n_i = 1.5 \times 10^{11}$ cm$^{-1}$) placed at a
distance `$d$' from the edge of the quantum well (i.e. from the
GaAs-AlGaAs interface) 
and a background 3D charged impurity density ($n_{iw}$) which is
distributed uniformly throughout the inside of the GaAs quantum well
(which has a well thickness of 300 \AA). 
Results presented in Fig.~\ref{fig1} involve no approximation other
than assuming a uniform random distribution of the quenched charged
impurities (2D distribution with a fixed 2D impurity density $n_i$ for
the remote impurities placed at a distance $d$ from the quantum well
and 3D distribution with a variable 3D impurity density $n_{iw}$ for
the unintentional background impurities inside quantum well) as we
include the full  
quantitative effect of the quasi-2D nature of the quantum well width
in the calculation and calculate all the integrals [Eqs.~(\ref{eq1})
and (\ref{eq2})] for $\tau_t^{-1}$ and $\tau_q^{-1}$ numerically
exactly. Of course, our basic theory is a leading-order theory in the
impurity scattering strength which should be an excellent
approximation at the high carrier density of interest in the current
work where our focus is on very low-disorder and high-quality 2D
semiconductor systems. The use of the realistic 3D background impurity
distribution is easily reconciled with our minimal model in section II
by using: $n_i=n_R$,
$d_R = d +a/2$ and $n_B \approx n_{iw} a$, where $a$ ($=300$ \AA \; in
Fig.~\ref{fig1}) is the quantum well-width. 

Results in Fig.~\ref{fig1} are quite revealing of the physics
discussed already in section II. First, we  clearly see in
Fig.~\ref{fig1}(a) the trend that for large (small) $d$, the mobility
is determined by the background (remote) impurities, and hence for
$d=800$ ($100$) \AA, the mobility depends strongly (weakly) on the
background impurity density (until it becomes very large, leading to
$\mu < 10^6$ cm$^2$/Vs which is no longer a high-quality
situation). In particular, for $d=800$ \AA, $\mu^{-1}$ ($\propto
\tau_t^{-1}$) $\propto n_{iw}$ approximately, implying that
$\tau_t^{-1}$ is dominated almost entirely (for $d=800$ \AA) by the
background impurities in the quantum well. By contrast, for $d=100$
\AA, $\mu$ is almost independent of $n_{iw}$ for $n_{iw} \alt 10^{15}$
cm$^{-3}$ (corresponds to $n_B \sim 3 \times 10^9$ cm$^{-2}$),
indicating that $\tau_t^{-1}$ is dominated almost entirely by the
``remote" dopant scattering. In Fig.~\ref{fig1}(a)   
 we focus on the interesting $d=800$ \AA \; situation where the
 2-impurity model might apply -- obviously, for $d=100$ \AA, the
 remote dopants dominate both $\tau_t^{-1}$ and $\tau_q^{-1}$
 rendering the 2-impurity model inapplicable since $k_F d < 1$ for
 both ``remote" and ``background" impurities for small `$d$'.

\begin{figure}[t]
	\centering
	\includegraphics[width=1.\columnwidth]{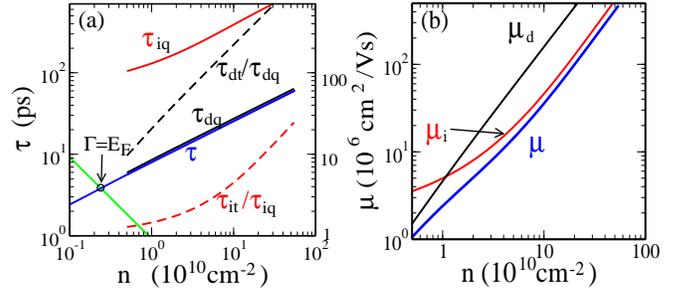}
	\caption{
(a) Calculated scattering times and (b)  mobilities as a function of carrier density. 
Here $\tau_{iq}$, $\mu_i$ ($\tau_{dq}$, $\mu_d$) indicate the single
particle relaxation 
time and mobility due to interface impurities at $d=0$ (remote impurities at
finite $d=800$ \AA), respectively, and $\tau 
=(\tau_{iq}^{-1} + \tau_{dq}^{-1})^{-1}$, $\mu =(\mu_i^{-1}+\mu_d^{-1})^{-1}$.
The Green line indicates $\tau_q = 0.92/n$ in units of ps
with $n$ measured by $10^{10}cm^{-2}$. The crossing point between green line and
blue line represents $\Gamma=E_F$.
The following parameters are used: $n_i (d=0) = 10^8 cm^{-2}$,
$n_d(d=800\AA) =10^{10}cm^{-2}$ and quantum well width $a=300$ \AA. At a density $n=10^{11}cm^{-2}$ we have
$\mu=35.7\times 10^6 cm^2/Vs$ and $\tau_q=25.5$ ps. The critical
density (i.e. the crossing point) is $n_c=0.24 \times 10^{10}cm^{-2}$.
}
\label{fig2}
\end{figure}

 In Fig.~\ref{fig1}(b) we show as a function of background disorder
 the calculated $\tau_q$ (as well as $\tau_t$) for $d = 800$ \AA, and
 it is clear that for $5\times10^{13} cm^{-3} < n_{iw} < 2 \times
 10^{15} cm^{-3}$ (i.e., over a factor of 40 increase in the
 background disorder!) $\tau_q$ (i.e., quality) remains almost a
 constant whereas $\tau_t$ (i.e., mobility) decreases approximately by
 a factor of 40 in this regime. Combining Figs.~\ref{fig1}(a) and (b),
 we  
then conclude that there could be an infinite series of samples, where
the mobility decreases from $\sim 20 - 40 \times 10^6$ cm$^2$/Vs to
below $10^6$ cm$^2$/Vs as $n_{iw}$ increases from $5\times 10^{13}$
cm$^{-3}$ to $2\times 10^{15}$ cm$^{-3}$, with all of them having
essentially the same quality as characterized by the quantum
scattering time $\tau_q \sim 3$ ps, corresponding to a quantum level
broadening of $\Gamma \sim \hbar/2\tau_q \sim 0.1$ meV. Results
shown in Fig.~\ref{fig1}, which are completely realistic, clearly
bring out the fact that,  when the underlying disorder has a basic
2-impurity model structure (one type of impurity with $k_Fd \gg1$ and
the other type with $k_F d \ll 1$), mobility and quality of individual
samples may very well be completely independent quantities.

We believe that the results of Fig.~\ref{fig1} completely
qualitatively (perhaps even quantitatively) explain the recent
``puzzling" finding\cite{gamez} that
the fragile 5/2 fractional quantum Hall effect (FQHE), which is
traditionally only studied in the highest quality samples with $\mu >
10^7$ cm$^2$/Vs, can actually be observed in much lower mobility
samples with $\mu \sim 10^6$ cm$^2$/Vs since, under suitable
circumstances (as in the results of Fig.~\ref{fig1}), it is possible
for samples with orders of magnitude different mobilities (i.e. values
of $\tau_t$) to have more  
or less the same ``quality" (i.e., the same value of $\tau_q$).

\begin{figure}[t]
	\centering
	\includegraphics[width=1.\columnwidth]{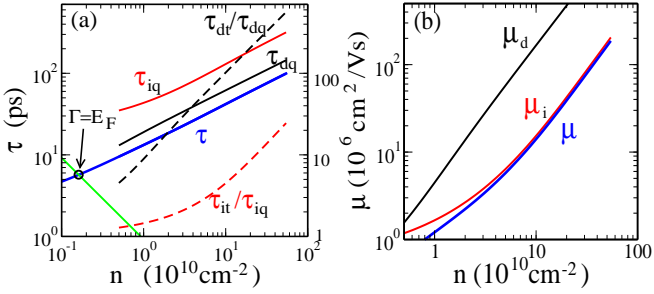}
	\caption{
The same as Fig.~\ref{fig2} with following parameters: $n_i (d=0) = 3\times
10^8 cm^{-2}$,  $n_d(d=500\AA) =3\times 10^{9}cm^{-2}$, and quantum well width $a=300$ \AA. 
At a density $n=10^{11}cm^{-2}$ we have
$\mu=14.1\times 10^6 cm^{2}/Vs$ and $\tau_q=42$ ps. The critical
density $n_c=0.161 \times 10^{10} cm^{-2}$.
}
\label{fig3}
\end{figure}

\begin{figure}[t]
	\centering
	\includegraphics[width=1.\columnwidth]{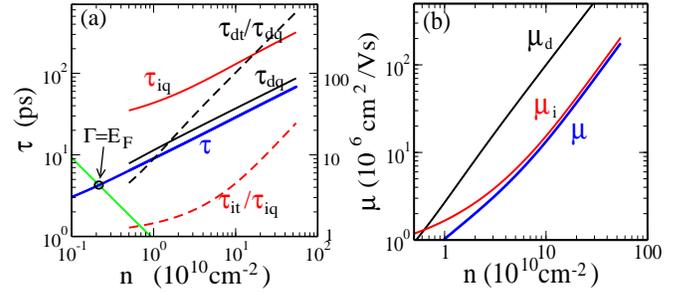}
	\caption{(a)
The same as Fig.~\ref{fig2} with following parameters: $n_i (d=0) = 3\times
10^8 cm^{-2}$, 
$n_d(d=500\AA) =5\times 10^{9}cm^{-2}$, and quantum well width $a=300$ \AA.
At a density $n=10^{11}cm^{-2}$ we have
$\mu=13.3\times 10^6 cm^{2}/Vs$ and $\tau_q=28.8$ ps. The critical
density $n_c =0.215 \times 10^{10}cm^{-2}$.
}
\label{fig4}
\end{figure}

\begin{figure}[t]
	\centering
	\includegraphics[width=1.\columnwidth]{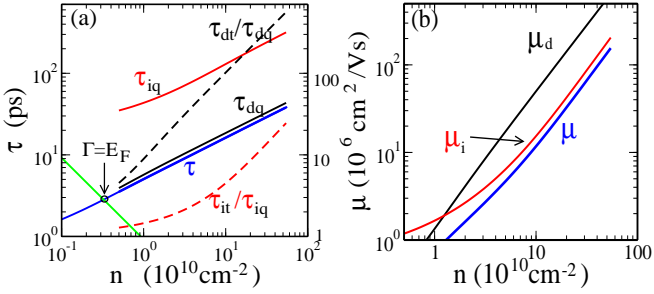}
	\caption{(a)
The same as Fig.~\ref{fig2} with following parameters: $n_i (d=0) = 3\times
10^8 cm^{-2}$, 
$n_d(d=500\AA) =10^{10}cm^{-2}$, and quantum well width $a=300$ \AA.
At a density $n=10^{11}cm^{-2}$ we have
$\mu=11.7\times 10^6 cm^{2}/Vs$ and $\tau_q=16.2$ ps. The critical
density $n_c =0.326 \times 10^{10}cm^{-2}$.
}
\label{fig5}
\end{figure}

In Figs.~\ref{fig2} -- \ref{fig5}, we make the above issue very clear
by showing realistic transport calculation results (for both
$\tau_t$ and $\tau_q$) in various situations within the 2-impurity
model. In each case, the high carrier density mobility is determined
by the background impurity scattering whereas the quality, i.e., the
quantum lifetime $\tau_q$ (or equivalently the single-particle level
broadening $\Gamma \sim \tau_q^{-1}$) is determined by remote impurity
scattering, creating a clear dichotomy where mobility and quality are
disconnected and the high-density mobility by itself does not provide
a unique characterization of the sample quality.

To make the physical implication of the mobility/quality dichotomy
very explicit, we have shown in each figure the carrier density where
the Ioffe-Regel criterion for strong localization, $\Gamma = E_F$, is
satisfied in each of these samples.\cite{IRC} (We mention that Figs.~\ref{fig2}
-- \ref{fig5} should be thought of as representing five distinct 2D
samples with fixed bare disorder each as described in each figure
caption, but with variable 2D carrier density, as for example, can be
implemented experimentally using an external back gate.) This $\Gamma
= E_F$ Ioffe-Regel point should be thought of as the critical density
below (above)  
which the system behaves insulating (metallic) as has recently been
discussed by us in details elsewhere \cite{IRC}. Such an apparent
disorder driven effective 2D metal-insulator transition (2D MIT) has been
extensively studied in the literature \cite{review}, and is usually
discussed in terms of the maximum mobility of the sample at high
carrier density.

One can think of the Ioffe-Regel criterion induced critical density
$n_c$ to be an approximate quantitative measure of the ``sample
quality" with $n_c$ decreasing (increasing) as the quality
improves. The corresponding approximate measure of the sample mobility
has traditionally been the so-called ``maximum mobility ($\mu_m$)", or
equivalently for high-mobility GaAs-based modulation-doped structures,
the measured mobility at the highest possible carrier density (since
for modulation-doped high-mobility structures, as can be seen in
Figs.~\ref{fig2} -- \ref{fig5} and as has been extensively
experimentally observed over the last 20 years, the sample mobility
decreases with decreasing carrier density and the typically quoted
sample mobility is always the one measured at the highest carrier
density). The n\"{a}ive expectation is that higher (lower) the maximum
mobility, lower (higher) should be the critical density for 2D MIT
since the sample quality should improve with sample mobility. 

\begin{table}[b]
\begin{tabular}{c | c | c}
\hline \hline
 Figure & $\mu_m $ ($10^6$ cm$^2$/Vs)  &  $ n_c$ ($10^{10}$ cm$^{-2}$)   \\
 \hline
 Fig. \ref{fig2} & 35.7  &   0.24    \\
 Fig. \ref{fig3} & 14.1  &    0.16    \\
 Fig. \ref{fig4} &  13.3  &    0.22   \\
 Fig. \ref{fig5} &  11.7  &     0.33 \\
\hline
\end{tabular}
\caption{$\mu_m$ is the mobility calculated at $n=10^{11}$ cm$^{-2}$ and $n_c $ represents the critical density calculated from  $\Gamma = E_F$.}
\end{table}

Using $\mu_m$ to be the mobility at $n=10^{11}$ cm$^{-2}$ in
Figs.~\ref{fig2} -- \ref{fig5}, we conclude that the sample quality,
if it is indeed determined entirely by the maximum mobility (or the
mobility at a very high carrier density), should decrease
monotonically as we go from the sample of Fig.~\ref{fig2} ($\mu_m =
35.7 \times 10^6$ cm$^{-2}$/Vs) to that of Fig.~\ref{fig5} ($\mu_m =
11.7\times 10^6$ cm$^{-2}$/Vs). We list below in Table I the
calculated critical density for each sample in Figs.~\ref{fig2} --
\ref{fig5} noting also the mobility $\mu_m$ at $n=10^{11}$ cm$^{-2}$.

We conclude from Table I that there is simply no one-to-one
relationship between mobility and quality in these numerical transport
results based on the 2-impurity model. For example, although the
``lowest mobility" sample (Fig.~\ref{fig5}, $\mu_m=11.7\times10^6$
cm$^2$/Vs) does indeed have the ``lowest quality" as reflected in the
highest value of $n_c$ ($\sim 0.33 \times 10^{10}$ cm$^{-2}$), the
highest mobility sample (with almost three times the mobility of all the
other samples) has the second highest value of $n_c$ ($\sim 0.24
\times 10^{11}$ cm$^{-2}$) instead of having the lowest $n_c$ as it
would if quality is determined exclusively by mobility. The other two
intermediate mobility samples with mobilities $14.1\times 10^6$
cm$^{-2}$/Vs and $13.3\times 10^6$ cm$^{-2}$/Vs also have their $n_c$
values ``reversed" ($0.16\times 10^{10}$ cm$^{-2}$ and $0.22 \times
10^{10}$ cm$^{-2}$, respectively) compared with what they should be if
the mobility really determined quality. We note that the samples of
Figs.~\ref{fig2} and \ref{fig4} have almost identical quality (i.e.,
essentially the same values of $n_c$) although the sample of
Fig.~\ref{fig2} has almost three times the high-density mobility as
that of Fig.~\ref{fig4}!

We do mention that the values of $n_c$ ($\sim 2 \times 10^9$
cm$^{-2}$) we obtain in our Figs.~\ref{fig2} -- \ref{fig5} are
consistent with the observed 2D MIT critical density in
ultra-high-mobility 2D GaAs structures where $n_c \sim 2 \times 10^9$
cm$^{-2}$ has been reported for $\mu_m \sim 10^7$ cm$^2$/Vs for $n
\alt 10^{11}$ cm$^{-2}$. \cite{lilly} 

The last set of numerical results we show for the 2-impurity model is
based on HIGFET
(heterojunction-insulator-gated-field-effect-transistor) structures
(in contrast to MODFET structures or
modulation-doped-field-effect-transistors, which we have discussed so
far in this paper) and is motivated by the recent experimental work by
Pan and his collaborators on the effect of disorder on the
observation, existence, and stability of the 5/2 FQHE in
high-mobility GaAs-AlGaAs HIGFET structures \cite{panprl2011}. 
This work of Pan {\it et al}. is closely related to similar work by
Gamez and Muraki \cite{gamez} and by Samkharadze {\it et al.}
\cite{samkharadze} who also studied disorder effects on the stability
of the 5/2 FQHE in modulation doped GaAs-AlGaAs 2D systems. All three
of these experimental studies conclude, using different phenomenology
and methodology, that the quality of the observed 5/2 FQHE in 2D
systems is not directly connected in an one-to-one manner with the
sample mobility, and it is possible to find robust 5/2 FQHE in samples
with mobility in the $\agt 10^6$ cm$^2$/Vs range whereas much of the
earlier work \cite{choi,dean} had to use ultra-high mobility ($ >
10^7$ cm$^2$/Vs) for the observation of stable 5/2 FQHE. This
observation by these three experimental groups of the mobility/quality
dichotomy is very similar to the theory being developed in the current
work with the only difference being that our work specifically focuses
on the quality being associated with the single-particle quantum
scattering rate $\tau_q^{-1}$ or the collisional level-broadening
$\Gamma \sim \tau_q^{-1}$ rather than the 5/2 FQHE gap since we do not
know of any quantitative microscopic theory which directly connects
FQHE gap values with disorder. We comment further on this feature
below in our discussion after presenting our HIGFET numerical
results.

A HIGFET system is different from modulation-doped quantum well
structures we considered so far in our work with the important
qualitative difference being that HIGFETs are undoped (except, of
course, for unintentional background charged impurities as represented
by $n_B$, which are unavoidable in a semiconductor) with no remote
modulation doping layer present in the system. Instead, the 2D
carriers are induced in the GaAs surface layer at the AlGaAs-GaAs
interface by a remote heavily doped gate placed very far from the
GaAs-AlGaAs interface. \cite{kane} Thus, HIGFETs are basically the
GaAs version of Si-MOSFETs (metal-oxide-semiconductor-field-effect
transistors) with the insulator being the AlGaAs layer instead of
SiO$_2$. An additional difference between 
HIGFETs and modulation-doped quantum wells is that the quasi-2D
carrier confinement in the HIGFET is in an asymmetric triangular
potential well (similar to MOSFETs \cite{andormp}) in contrast to the
symmetric square well confinement in the AlGaAs-GaAs quantum well
system. 

\begin{figure}[t]
	\centering
		\includegraphics[width=0.70\columnwidth]{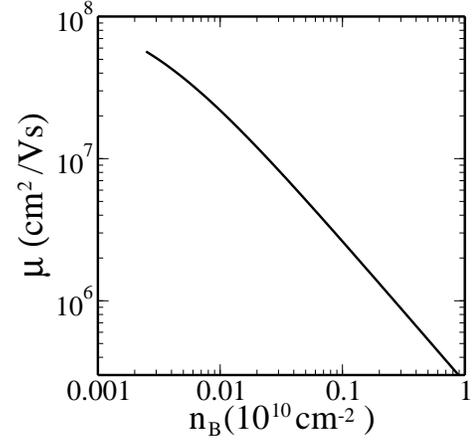}
	\caption{
(a) The calculated mobility as a function of background impurity density $n_B$ in a GaAs HIGFET structure.
Here $n_R=10^{13}$ cm$^{-2}$, $d_R=630$ nm, and the carrier density $n=4.7 \times 10^{11}$ cm$^{-2}$ are used.
}
\label{fig6}
\end{figure}

Given that HIGFETs have no intentional modulation doping, it may
appear that the 2-impurity model is simply inapplicable here since the
background unintentional charged impurities seem to be the only
possible type of Coulomb disorder in the system so that the system
should belong to a 1-impurity disorder description (i.e., just the
unintentional background random charged impurities). This is, however,
incorrect because the presence of the far-away gate, which induces the
2D carriers  
in the HIGFET, introduces remote charged disorder (albeit at a very
large value of $d$) arising from the gate charges which must be
present due to the requirement of charge neutrality. We, therefore,
use exactly the same minimal 2-impurity model for the HIGFETs that we
have used for the modulation doped systems assuming $n_R$ to be the
charged impurity density on the far away gate at a very large distance
$d_R$ away from the induced 2D electron layer on the GaAs side of the
GaAs-ALGaAs interface. (Later in this section we will present results
for a realistic 3-impurity model in order to provide a quantitative
comparison with the HIGFET data of Pan {\it et
  al}. \cite{panprl2011}.)

\begin{figure}[t]
	\centering
	\includegraphics[width=1.0\columnwidth]{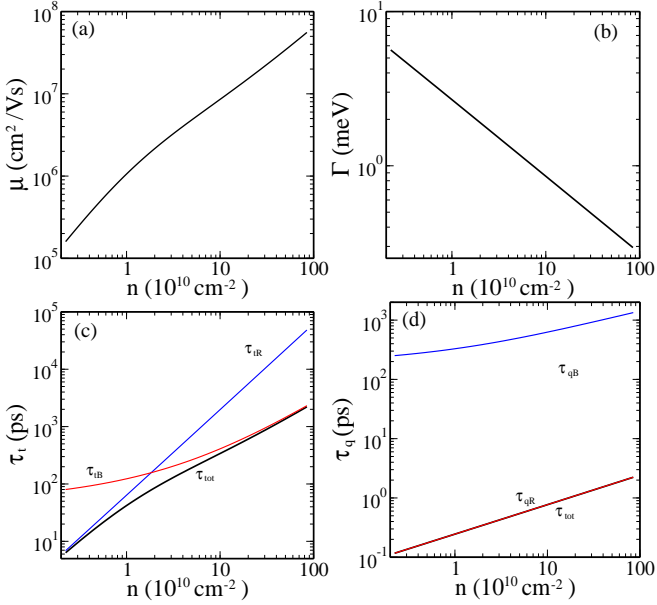}
	\caption{
The calculated (a) mobility, $\mu$,  (b) level broadening, $\Gamma$,
(c) transport lifetime, $\tau_t$ and (d) quantum lifetime, $\tau_q$,
as a function of carrier density in a GaAs HIGFET structure with
parameters  $n_R=10^{13}$ cm$^{-2}$, $d_R=630$ nm, and $n_B=1.69
\times 10^8$ cm$^{-2}$. 
}
\label{fig7}
\end{figure}

\begin{figure}[t]
	\centering
	\includegraphics[width=1.0\columnwidth]{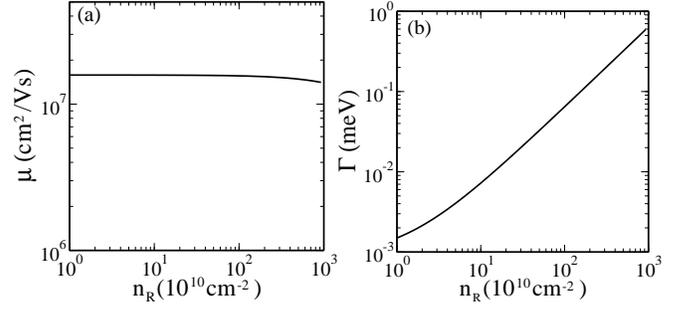}
	\caption{
The calculated (a) mobility and (b) level broadening as a function of
remote impurity density in a GaAs HIGFET structure with parameters
$n=1.8 \times 10^{11}$ cm$^{-2}$, $d_R=630$ nm, and $n_B=1.69 \times
10^{10}$ cm$^{-2}$. 
}
\label{fig8}
\end{figure}

In Figs.~\ref{fig6} -- \ref{fig8} we show our full numerical results
of a n-GaAs HIGFET structure using the 2-impurity model. The specific
HIGFET structure used for our numerical calculations is motivated by
the sample used in Ref.~[\onlinecite{panprl2011}], but we do not
attempt any quantitative comparison with the experimental transport
results, which necessitates a 3-impurity model to be described
later. At this stage, i.e., for Figs.~\ref{fig6} -- \ref{fig8}, our
goal is to establish the mobility/quality dichotomy for HIGFET 2D
systems based on our minimal 2-impurity model.

In Fig.~\ref{fig6} we show the calculated mobility as a function of
$n_B$, the background impurity density and in Fig.~\ref{fig7} we show
the mobility ($\mu$), the level broadening ($\Gamma$), the transport
lifetime ($\tau_t$), and the quantum lifetime ($\tau_q$) as a function
of the 2D carrier density  
$n$ in a GaAs HIGFET structure using the 2-impurity model with $n_R =
10^{13}$ cm$^{-2}$; $d_R = 630$ nm; $n_B = 1.69 \times 10^8$
cm$^{-2}$. We note that 
$\mu  =   ne\tau_t$ and $\Gamma  =  \hbar/2\tau_q$
are simple measures of mobility and quality which are directly
linearly connected to $\tau_t$ and $\tau_q^{-1}$. Our choice of $d_R =
630$ nm is specifically aimed at the sample of
Ref.~[\onlinecite{panprl2011}] where the gate is located 630 nm away
from the GaAs-AlGaAs interface. Our choice of $n_R=10^{13}$ cm$^{-2}$
and $n_B=1.69 \times 10^8$ cm$^{-2}$ is arbitrary at this stage (and
the  precise choice here is irrelevant with respect to our qualitative
conclusions) except that this combination of a large (small) $n_R$
($n_B$) is the appropriate physical situation in high-quality  
HIGFETs. Our choice of $n_R$, $n_B$, and $d_R$ (which we get from the
actual experimental system) gives the correct 2D ``maximum" mobility
of $\mu = 14 \times 10^6$ cm$^2$/Vs at a 2D carrier density of $n=4.7
\times 10^{11}$ cm$^{-2}$ consistent with the experimental sample in
Ref.~[\onlinecite{panprl2011}] as shown in Fig.~\ref{fig6}. 

The calculated mobility in Fig.~\ref{fig6} decreases monotonically
with increasing $n_B$, and we choose $n_B = 1.69 \times 10^8$
cm$^{-2}$ to get the correct maximum mobility of $14 \times 10^6$
cm$^2$/Vs reported in Ref.~[\onlinecite{panprl2011}] with the
corresponding value of the level broadening being 0.638 meV at the
same density. We emphasize that the level broadening (or equivalently,
$\tau_q$) here is determined entirely by the remote scattering from
the gate in spite of the gate being an almost macroscopic distance
($\sim 0.6 \mu m$) away from the 2D electrons -- changing $n_B$ by
even a factor of 100 does not change the the value of $\Gamma$ or
$\tau_q^{-1}$ (but does change mobility $\mu$ or $\tau_t^{-1}$ by a factor of
100) whereas the mobility is determined entirely by the background
scattering (and therefore changing $n_R$ does not affect the
mobility).

In Fig.~\ref{fig7} we show the calculated $\mu$, $\Gamma$, $\tau_t$,
and $\tau_q$ (remembering $\mu = ne \tau_t$ and $\Gamma =
\hbar/2\tau_q$) as a  function of 2D carrier density $n$ for fixed
$n_R$, $d_R$, $n_B$ as shown. These results clearly show the
mobility/quality dichotomy operational within the 2-impurity model in
this particular HIGFET structure (for the chosen realistic disorder
parameters $n_R$, $d_R$, $n_B$).  
For $n \agt 2\times 10^{10}$ cm$^{-2}$, the mobility is determined
essentially by the background impurity scattering (i.e. $n_B$) whereas
the level broadening or the quantum scattering rate is determined
entirely by the remote scattering for the entire density range ($10^9
cm^{-2} < n < 10^{12} cm^{-2}$) shown in Fig.~\ref{fig7}. At low
carrier density ($n \alt 10^{10}$ cm$^{-2}$) $k_Fd_R$ ($\alt 10$) is
no longer very large, and given the rather large value of $n_R$
($=10^{13}$ cm$^{-2}$) corresponding to the remote gate charges, the
scattering by $n_R$ starts affecting the mobility. But, the high
density mobility (determined by $n_B$) and the quality at all density
(determined by $n_R$) are still completely independent quantities, and
therefore it is possible for the quality (e.g., the FQHE gap at high
density) to be completely independent of the mobility as found
experimentally in Ref.~[\onlinecite{panprl2011}].

\begin{figure}[t]
	\centering
		\includegraphics[width=0.70\columnwidth]{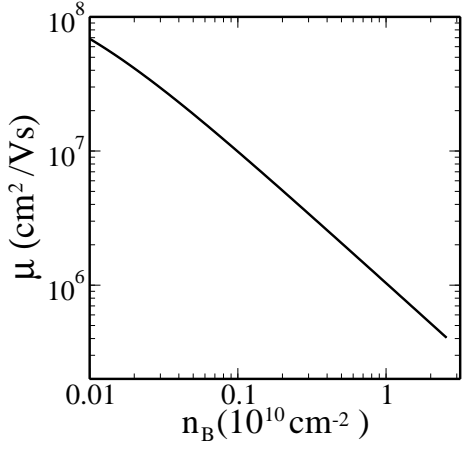}
	\caption{
(a) The calculated mobility as a function of background impurity
          density $n_B$ in a GaAs MODFET structure with a well width
          $a=300$\AA. 
Here $n=n_R=1.8\times 10^{11}$ cm$^{-2}$ and $d_R=2000$ nm are used.
}
\label{fig9}
\end{figure}

This is demonstrated explicitly in Fig.~\ref{fig8} where we show that
the variation in the mobility is essentially non-existent for four
orders of magnitude changes in $n_R$ whereas $\Gamma$ changes
essentially by four orders of magnitude. Similarly, Fig.~\ref{fig7}
indicates that (since $\tau_{qB}^{-1} \propto \Gamma \propto n_B$), a
2 orders of magnitude change in $n_B$ will hardly change $\Gamma$, but
$\mu$ will change by 2 orders of magnitude (due to a 2-orders of
magnitude change in $n_B$) at high carrier density. 

\begin{figure}[t]
	\centering
	\includegraphics[width=1.0\columnwidth]{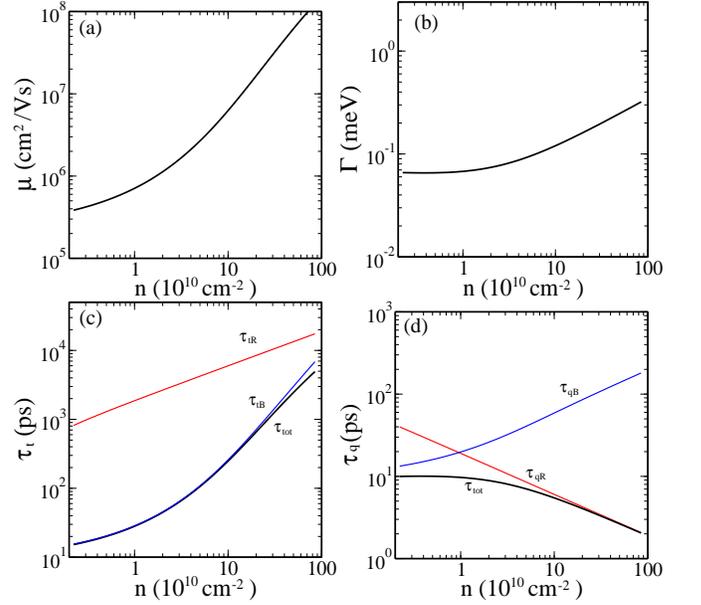}
	\caption{
The calculated (a) mobility, $\mu$,  (b) level broadening, $\Gamma$,
(c) transport lifetime, $\tau_t$ and (d) quantum lifetime, $\tau_q$,
as a function of carrier density in a GaAs MODFET structure with a
well width $a=300$ \AA. Here the parameters  $n_R=n$, $d_R=200$ nm,
and $n_B=6.8 \times 10^8$ cm$^{-2}$ are used. 
}
\label{fig10}
\end{figure}

\begin{figure}[t]
	\centering
	\includegraphics[width=1.0\columnwidth]{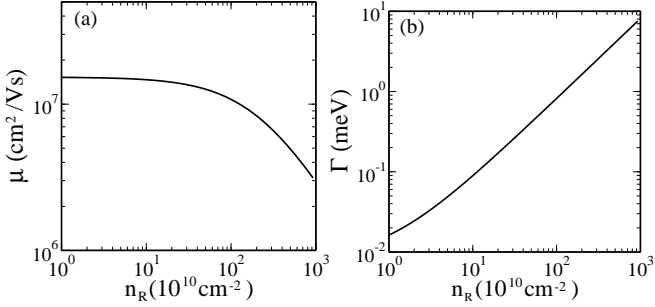}
	\caption{
The calculated (a) mobility and (b) level broadening as a function of
remote impurity density in a GaAs MODFET structure with a well width
$a=300$ \AA. The parameters $n=1.8 \times 10^{11}$ cm$^{-2}$,
$d_R=200$ nm, and $n_B=6.8 \times 10^{10}$ cm$^{-2}$ are used. 
}
\label{fig11}
\end{figure}

We believe that our Figs.~\ref{fig6} -- \ref{fig8} provide a complete
explanation for the puzzling observation in
Ref.~[\onlinecite{panprl2011}] where a drop in the mobility of the
sample at high carrier density hardly affected its quality as
reflected in the measured 5/2 FQHE energy gap. This is because the
high carrier density mobility is determined by background impurity
density $n_B$ which does not affect the quality at all whereas the
quality is affected by remote scattering which does not much affect
the mobility at high carrier density. For the sake of completeness,
and to make connection with the interesting recent works of
Refs. [\onlinecite{gamez,samkharadze}], who also independently
conclude in agreement with Pan {\it et al}.\cite{panprl2011} that very
high mobility ($>10^7$ cm$^2$/Vs) is not necessarily required for the
experimental observation of a robust 5/2 FQHE in standard
modulation-doped GaAs quantum wells (in contrast to Pan's usage of
undoped HIGFETs), we show in Figs.~\ref{fig9} -- \ref{fig11} ( which
correspond to the HIGFET results shown in Figs.~\ref{fig6} --
\ref{fig8} respectively) our calculated transport results for a
modulation-doped quantum well structure with a high-density mobility
identical (i.e., $14\times 10^6$ cm$^2$/Vs) to the HIGFET structure
considered in Figs.~\ref{fig6} -- \ref{fig8}.

The main differences between the 2D systems for Figs.~\ref{fig6} --
\ref{fig8} (HIGFET) and Figs.~\ref{fig9} -- \ref{fig11} (MODFET) are
the following:  
(1) the HIGFET has a triangular quasi-2D confinement potential
(determined self-consistently by the carrier density) and the MODFET
has a square-well confinement imposed by the MBE-grown
AlGaAs-GaAs-AlGaAs structure with a given confinement width ($a=30$ nm
in Figs.~\ref{fig9} -- \ref{fig11}); (2) the 2D carriers are induced
by a very far away gate in the HIGFET whereas it is induced by the
remote dopants (we choose $n_R=n$ in Figs.~\ref{fig9} -- \ref{fig11})
in the modulation doping layer (we choose $d_R=200$ nm in
Figs.~\ref{fig9} -- \ref{fig11}); (3) the specific necessary values of
$n_B$ are somewhat different in the two systems in order to produce
the same high-density maximum mobility. 
The quantitative differences described in items (1) - (3) above are
sufficient to produce substantial differences between the numerical
results in the HIGFET and the MODFET system as can easily be seen by
comparing the results of Figs.~\ref{fig6} -- \ref{fig8} with those of
Figs.~\ref{fig9} -- \ref{fig11}, respectively, although we ensured
that both have exactly the same high-density mobility ($\mu_m = 1.4
\times 10^7$ cm$^2$/Vs). However, qualitatively the two sets of
results shown in Figs.~\ref{fig6} -- \ref{fig8}  
and \ref{fig9} -- \ref{fig11} are similar in that the mobility
(quality) at high carrier density ($>10^{10}$ cm$^{-2}$) is invariably
determined by the background (remote) scattering respectively, leading
to the possibility that a substantial change in mobility (quality) by
changing $n_B$ ($n_R$) respectively may not at all affect quality
(mobility), and thus it is possible at high carrier density for a
system to have a modest mobility ($\sim 10^6$ cm$^2$/Vs) by having a
large $n_B$ with little adverse effect on quality (i.e.,
$\Gamma$). Thus, the experimental observations in
Refs.~[\onlinecite{gamez,panprl2011,samkharadze}] are all consistent
with our theoretical results.

\begin{figure}[t]
	\centering
	\includegraphics[width=1.0\columnwidth]{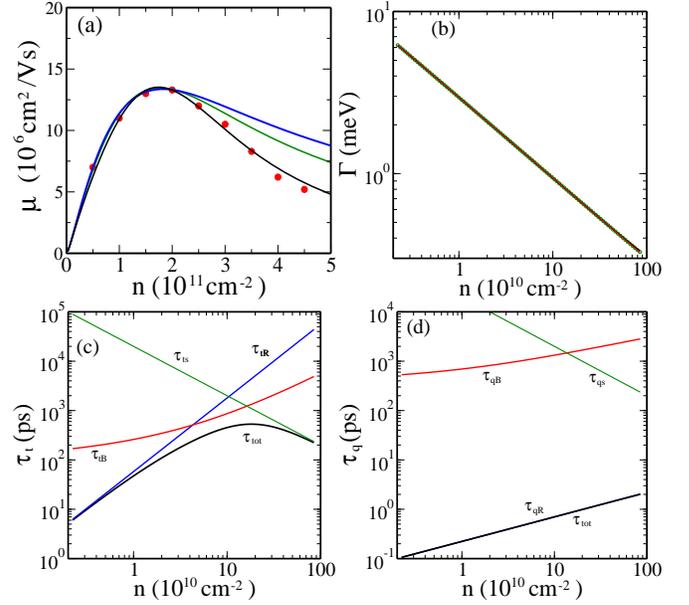}
	\caption{
The calculated (a) mobility and (b) level broadening with long range
          remote impurity at $d_R$, short range impurities at the
          interface, and background short range impurities. We assume that
          the density dependence of scattering time with interface
          short range impurities is $\tau_{qs}^{-1} \propto n^{\alpha}$. In (a)
          the blue line is calculated with $n_B = 0.8 \times 10^8
          cm^{-2}$ and $\tau_{qs}^{-1} = 5\times 10^7 n$. The green line
          is calculated with  $n_B = 0.9 \times 10^8
          cm^{-2}$ and $\tau_{qs}^{-1} = 1.85\times 10^7 n^{1.3}$. The black line
is calculated with $n_B = 1.08 \times 10^8
          cm^{-2}$ and $\tau_{qs}^{-1} = 0.19\times 10^7 n^2$. The red dots
          are experimental data from Pan {\it et al.}\cite{panprl2011}
In (c) and (d) we show the total scattering times as well as the
individual scattering 
          time corresponding to the each scattering source. Here $n_B
          = 0.8 \times 10^8 
          cm^{-2}$ and $\tau_{qs}^{-1} = 5\times 10^7 n$ are used. 
}
\label{fig12}
\end{figure}

\begin{figure}[t]
	\centering
	\includegraphics[width=1.0\columnwidth]{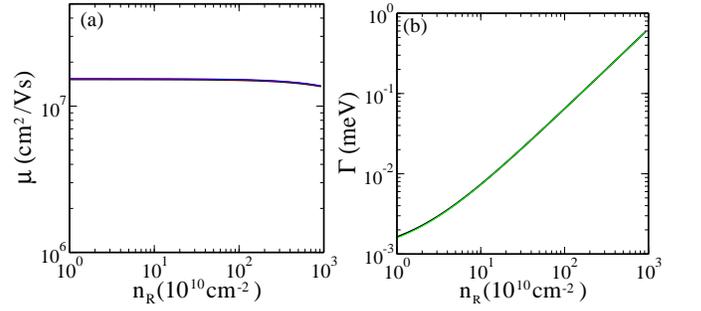}
	\caption{
The calculated (a) mobility and (b) level broadening as a function of the
 remote impurity density. The same parameters as Fig.~\ref{fig12}(a) are used. 
}          
\label{fig13}
\end{figure}

Finally, we show in Figs.~\ref{fig12} and \ref{fig13} the numerical
transport results for the HIGFET structure (of Figs.~\ref{fig6} --
\ref{fig8}) using a more realistic 3-impurity model going beyond the
2-impurity model mostly used in our current work. The 3-impurity model
is necessary for obtaining agreement between experiment
\cite{panprl2011} and theory since experimentally the measured
mobility, $\mu(n)$, as a function of carrier density manifests
non-monotonicity with a maximum in the mobility around $n \sim 2\times
10^{11}$ cm$^{-2}$. Such a non-monotonicity, where $\mu$ increases
(decreases) with increasing $n$ at low (high) carrier density, is
common in Si-MOSFETs \cite{andormp}, and is known to arise from
short-range interface scattering which becomes stronger with
increasing carrier density as the self-consistent confinement of the
2D carriers becomes stronger and narrower pushing the electrons close
to the interface and thus increasing the short-range interface
roughness scattering as well as the alloy disorder scattering in
AlGaAs as the confining wave function tail of the 2D electrons on the
GaAs side pushes into the Al$_x$Ga$_{1-x}$As side of the barrier. We
include this realistic short-range scattering effect, which becomes
important 
at higher carrier density leading to a decrease of the mobility at
high density (as can be seen in Fig.~\ref{fig12}(a)). Importantly,
however, this higher-density suppression of mobility 
(by a factor of 3 in Fig.~\ref{fig12}(a) consistent with the
observation of Pan {\it et al}. \cite{panprl2011}) has absolutely no
effect on the quality (see Fig.~\ref{fig12}(b)) with the level
broadening $\Gamma$ decreasing monotonically with increasing carrier
density (since $\Gamma$ is determined essentially entirely by the
remote scattering -- see Fig.~\ref{fig12}(d)). Thus, we see an apparent
paradoxical situation (compare Figs.~\ref{fig12}(a) and (b)) where the
mobility decreases at higher carrier density, but the quality keeps on
improving with increasing carrier density! This is precisely the
phenomenon observed by Pan {\it et al}. \cite{panprl2011} who found
that, 
although the mobility itself decreased in their sample by a factor of
3 at higher density, the sample quality, as measured by the 5/2 FQHE
gap, improved with increasing density precisely as we predict in our
work. In Fig.~\ref{fig13} we show that the 3-impurity model, except
for allowing the mobility to decrease at high carrier density due to
the increasing dominance of short-range scattering (thus bringing
experiment and theory into agreement at high density in contrast to
the 2-impurity results), has no effect on the basic quality/mobility
dichotomy being discussed in this work -- for example,
Fig.~\ref{fig13} shows that while quality decreases (i.e., $\Gamma$
increases) monotonically with increasing remote scattering, nothing
basically happens to the mobility! 

Before concluding this section, we provide a critical and quantitative
theoretical discussion of two distinct experiments (one from 1993
\cite{duprl1993} and the other from 2011 \cite{samkharadze}),
separated by almost 20 years in time, involving high-mobility 2D
semiconductor structures in the context of the mobility versus quality
question being addressed in the current work. Our reason for focusing
on these two papers is because both report $\tau_t$ and $\tau_q$ for
the samples used in these experimental studies, thus enabling us to
apply our theoretical analyses quantitatively to these samples.

In ref.~[\onlinecite{duprl1993}], two GaAs-AlGaAs heterojunctions were
used (samples A and B) with the following characteristics
\cite{duprivate}: Sample A: $n=1.1 \times 10^{11}$ cm$^{-2}$; $\mu = 6.8
\times 10^6$ cm$^2$/Vs; $\tau_t=270$ ps; $\tau_q = 9$ ps, and
Sample B: $n=2.3 \times 10^{11}$ cm$^{-2}$; $\mu = 12
\times 10^6$ cm$^2$/Vs; $\tau_t=480$ ps; $\tau_q = 4.5$ ps.
Both samples A and B have the same setback distance of $d_R=80$ nm
for the remote dopants, but we should consider $d_A > d_B \agt 80$ nm
since sample A has a lower carrier density and therefore the quasi-2D
layer thickness for sample A must be slightly higher since the
self-consistent confinement potential must be weaker in A than in B
due to its lower density. We note that sample A and B indeed manifest
the mobility/quality dichotomy in that A (B) has higher (lower)
quality (i.e. $\tau_q$), but lower (higher) mobility!

We start by assuming the absence of any background impurity scattering
($n_B=0$), then the asymptotic formula for $k_F d_R \gg 1$ applies to
both samples, giving, $\tau_t \sim (k_Fd_R)^3/n_R$; $\tau_q \sim
(k_Fd_R)/n_R$. Making the usual assumption $n_R = n$, since no
independent information is available for $n_R$, we conclude that the
theory predicts $\tau_t^B/\tau_t^A = \sqrt{n_B/n_A} (d_B/d_A)^3
\approx 1.4$ assuming $d_B \approx d_A$, and $\tau_q^B/\tau_q^A =
\sqrt{n_A/n_B} d_B/d_A \approx 0.7$ assuming $d_B \approx
d_A$. Experimentally, A and B samples satisfy: $\tau_t^B/\tau_t^A
\approx 1.8$; $\tau_q^B/\tau_q^A =0.5$.
Thus, just the consideration of only remote dopant scattering which
must always be present is all modulation-doped samples already gives
semi-quantitative agreement between theory and experiment including an
explanation of the apparent paradoxical finding that the sample B with
higher mobility has a lower quality! The key here is that the higher
density of sample B leads to a higher mobility, but also leads to a
higher values of $\tau_q^{-1}$ (and hence lower quality) by virtue of
higher carrier density necessitating a higher value of $n_R$ leading
to a lower value of $\tau_q$ [see, for example, Fig.~\ref{fig01}(c)
  where increasing $n=n_R$ leads to increasing (decreasing) $\tau_t$
  ($\tau_q$)]. 

We can actually get essentially precise agreement between theory and
experiment for the dichotomy in samples A and B of
Ref.~[\onlinecite{duprl1993}], with $\tau_t$ higher (lower) in sample
B (A) and $\tau_q$ higher (lower) in sample A (B) by incorporating the
fact that a higher (by a factor of 2) carrier density in sample B
compared with sample A makes $d_A>d_B$ due to self-consistent
confinement effect in heterostructures and hence the theoretical
ratios of $\tau^A$ and $\tau^B$ change from the values given above to
$\tau_t^B/\tau_t^A <1.4$ and $\tau_q^B/\tau_q^A \approx 0.5$ (i.e.,
$<0.7$). This means that while the quality ratio $\tau_q^B/\tau_q^A$
of samples A and B can be understood quantitatively on the basis of
remote scattering (which determines the quality almost exclusively in
high-mobility modulation-doped structures), the mobility ratio
$\tau_t^B/\tau_t^A$ is not determined exclusively by remote dopant
scattering. Inclusion of somewhat stronger background disorder
scattering in sample A compared with sample B immediately gives
$\tau_t^B/\tau_t^A =1.8$ in agreement with experiment. Thus, we see as
asserted by us theoretically, quality and mobility are mainly
controlled  by distinct scattering mechanisms (quality by remote
scattering and mobility by background scattering) in the data of
ref.~[\onlinecite{duprl1993}] providing an explicit example of the
mobility/quality dichotomy as far back as in 1993 when this dichotomy was
not discussed at all in the literature.

Considering now the samples used in ref.~[\onlinecite{samkharadze}],
there are again two distinct modulation-doped quantum well samples
with the following sample specifications: Sample A: $a=56$ nm; $d_R =
320$ nm; $n=8.3 \times 10^{10}$ cm$^{-2}$; $\mu = 12\times 10^6$
cm$^2$/Vs; $\Gamma = 0.24$ K, and Sample B: $a=30$ nm; $d_R =
78$ nm; $n=2.78 \times 10^{11}$ cm$^{-2}$; $\mu = 11\times 10^6$
cm$^2$/Vs; $\Gamma = 2.04$ K. Thus, in this case \cite{samkharadze},
although the two samples have almost identical mobilities, the
lower-density sample A has almost 8 times higher quality with
$\Gamma_B/\Gamma_A = \tau_q^A/\tau_q^B \approx 8$. We note that the
lower quality sample has three times the carrier density, and going
back to our Figs.~\ref{fig01} -- \ref{fig04}, we see that a higher
carrier density $n$ ($=n_R$) always leads to higher mobility and lower 
quality since the quality (i.e., $\tau_q$)
is determined mostly by long-range remote scattering whereas the
mobility is determined by a combination of both remote and background
scattering with the background scattering often dominating the
mobility. The fact that $d_R^A \gg d_R^B$ considerably improves the
quality of sample A with respect to sample B without much affecting
the mobility since the quality (mobility) is limited by remote
(background) scattering.

Using the asymptotic formula (for $k_Fd \gg 1$), $\tau_q \propto k_F d
/n_R$ and $n_R \approx n$, we conclude for the comparative quality of
the two samples: $\tau_q^A/\tau_q^B = \Gamma_B/\Gamma_A \approx
\sqrt{n_B/n_A} d_A/d_B \approx 8$ where we use $d_A = 348$ nm and $d_B
= 93$ nm by taking into account their differences in both the set back
distances and the well thickness. Experimentally, $\Gamma_B/\Gamma_A
\approx 8.5$ in excellent agreement with the theoretical estimate. The
fact that the mobilities of A and B are similar is easily explained by
their similarity with respect to background disorder with sample B
having somewhat larger value of unintentional background impurity
density than sample A. Thus, our mobility/quality theoretical
dichotomy is in perfect accord with the data of
ref.~[\onlinecite{samkharadze}].

We now conclude this section by mentioning that we have used the
quantum lifetime (or the single particle scattering time) $\tau_q$ (or
equivalently $\Gamma \propto \tau_q^{-1}$) as a measure of the quality
because it is well-defined and theoretically
calculable. Experimentally, the quality can be defined in a number of
alternative ways as , for example, done in the recent experiments
\cite{panprl2011,gamez,samkharadze} where the 5/2 FQHE gap is used as
a measure of the quality. There is no precise microscopic theory for
calculating disorder effects on the FQHE gap, 
but there are strong indications \cite{dean,morf,samkharadze} that the
FQHE gap $\Delta_{\Gamma}$ in the presence of finite disorder scales
approximately as 
\begin{equation}
\Delta_{\Gamma} \approx \Delta_0 - \Gamma,
\label{eq48}
\end{equation}
where $\Gamma$ is indeed the quantum level broadening we use in our
current work as the measure of quality and $\Delta_0$ is the FQHE gap
in the absence of any disorder. If this is even approximately
true (as it seems to be on empirical grounds), then our current
theoretical work shows complete consistency 
with the recent experimental results concerning the dichotomy between
mobility and FQHE gap values in the presence of disorder. In this
context, it may be worthwhile to emphasize an often overlooked fact:
the mobility itself (i.e., $\tau_t^{-1}$ and {\it not} $\tau_q^{-1}$)
can be converted into an energy scale by writing (for GaAs) 
\begin{equation}
\Gamma_{\mu} = \frac{\hbar}{2\tau_t} \approx (10^{-4}/\tilde{\mu}) \;
      {\rm meV} \; \approx (.01/\tilde{\mu}) \; {\rm K}, 
\end{equation} 
where $\tilde{\mu} = \mu/(10^7 cm^2/Vs)$.
Thus, a mobility of $10^7$ cm$^2$/Vs corresponds only to a
broadening of 10 mK which is miniscule compared with the
theoretically calculated \cite{morf,morf2} 5/2 FQHE gap of $2-3$ K!
Even a mobility of $10^6$ cm$^2$/Vs corresponds to a mobility
broadening of only 100 mK, which is much less than the expected 5/2 FQHE
gap. Thus, the quality of the 5/2 FQHE cannot possibly be determined
directly by the mobility value (unless the mobility is well below
$10^6$ cm$^2$/Vs) and there must be some other factor controlling the
quality, which we take to be the quantum level broadening in this
work. It must be emphasized here that the mobility/quality dichotomy
obviously arises from the underlying disorder in high-mobility
semiconductor structures being long-ranged. If both mobility and
quality are dominated by short-range disorder, then $\tau_q \approx
\tau_t$, and a mobility of $10^6$ cm$^2$/Vs with $\Gamma \approx 100$
mK will be a high-quality sample!

In concluding this section, we should mention that the very first
experimental work we know of where the mobility/quality dichotomy was
demonstrated and noted explicitly in the context of FQHE physics is a
paper by Sajoto et al. \cite{sajorto} from the Princeton group which
appeared in print an astonishing 24 years ago!  In this work, (see the
``Note added in proof" in Ref.~[\onlinecite{sajorto}]), it was
specifically stated that the samples used by Sajoto et al. manifested
as strong FQHE states as those observed in other samples from other
groups with roughly $5-10$ times the mobility of the Sajoto et
al. samples, thus providing a clear and remarkable early example of
the mobility/quality dichotomy much discussed during the last couple
of years in the experimental 2D literature.  We note that the samples
used by Sajoto et al. \cite{sajorto} had unusually large set-back
distances ($d_R \sim 270$ nm), leading to rather small values of
$\tau_q^{-1}$ and $\Gamma$ corresponding to our theory although the
mobility itself, being limited by background impurity scattering
(i.e. by $n_B$), was rather poor ($\sim 10^6$ cm$^2$/Vs).  We believe
that the reason the samples of Sajoto et al. had such high quality in
spite of having rather modest mobility is the mobility/quality
dichotomy studied in our work where the mobility determined by
background scattering is disconnected from the quality determined by
the remote dopant scattering. 

\section{summary and conclusion}

In summary, we have theoretically discussed the important issue of
mobility versus quality in high-mobility 2D semiconductor systems such
as modulation-doped GaAs-AlGaAs quantum wells and GaAs undoped HIGFET
structures. We have established, both analytically (section II) and
numerically (section III), that modulation-doped (or gated) 2D systems
should generically manifest a mobility/quality dichotomy, as often
observed experimentally, due to the simple fact that mobility and
quality are often determined by different underlying disorder
mechanisms in 2D semiconductor structures -- in particular, we show
definitively that in many typical situations, the mobility (quality)
is controlled by near (far) quenched charged impurities, particularly
at higher carrier density and higher mobility samples. We
show that often the 2D mobility (or equivalently, the 2D transport
scattering time)  is controlled by the unintentional background
charged impurities in the 2D layer whereas the quality, which we have
parameterized throughout our work by the quantum single-particle
scattering time (or equivalently, the quantum level broadening), is
controlled by the remote charged impurities in the modulation doping layer whose
presence is necessary for inducing carriers in the 2D layer. Somewhat
surprisingly, we show that the same mobility/quality dichotomy could
actually apply to undoped HIGFET structures where the charges on the
far-away gate play the role of remote scattering mechanism. Quite
unexpectedly, we show that a very far away gate (located almost
$10^{-4}$ cm away from the 2D layer) can still completely dominate
the quantum level broadening, while at the same time having no effect
on the mobility. We develop a minimal 2-impurity model (near and far
or background and remote) which is sufficient to explain all the
observed experimental features of the mobility/quality dichotomy. The
key physical point here is that the dimensionless parameter `$k_F d$',
where $k_F \propto \sqrt{n}$ is the Fermi wave number of the 2D
electron system and `$d$' is the distance of the relevant charged
impurities from the 2D system completely controls the mobility/quality
dichotomy. Impurities with $k_F d \gg 1$ ($\ll 1$) could totally
dominate quality (mobility) without affecting the other property at
all. We give several examples of situations where identical or very
similar sample mobilities at high carrier density could lead to very
different sample qualities (i.e., quantum level broadening differing
by large factors) and vice versa. 
The mobility/quality dichotomy in our minimal 2-impurity model arises
from the exponential suppression of the large angle scattering by
remote charged impurities which leads to the interesting situation that remote
scattering contributes little to the resistivity, but a lot to the
level broadening through the accumulation of substantial small angle
scattering. 
We emphasize that the mobility/quality dichotomy arises entirely from
the long-range nature of the underlying disorder, and would disappear
completely if the dominant disorder in the system is short-ranged.

It is important to realize that $\tau_t$ (mobility) and $\tau_q$
(quality) both depend not only on the disorder strength, but also on
the carrier density, i.e., $\tau_{t,q} \equiv
\tau_{t,q}(n,n_R,d_R,n_B)$. Thus, even within the 2-impurity model
(parameterized by disorder parameters $n_R$, $d_R$, $n_B$),
$\tau_{t,q}$ are both functions of carrier density. For very low
carrier density, the dimensionless parameter $k_Fd$ may be small for
all relevant impurities in the system, and eventually our 2-impurity
model will then fail since at such a low carrier density, all
impurities are essentially near impurities with the distinction
between R-impurities and B-impurities being merely a semantic
distinction with no real difference. Mobility and quality at such low
densities then will behave similarly. The same situation may also
apply as a matter of principle at very high carrier densities (i.e.,
very large $k_F$) where all impurities may satisfy $k_F d \gg 1$ and
thus act as far impurities, again leading to a breakdown of the
2-impurity model. This density dependence of the 2-impurity model with
respect to mobility/quality dichotomy is, however, a non-issue for our
current work since (1) typically, samples are characterized by their
mobility values at some fixed high (but not too high) carrier density
($n \sim 10^{11} - 4 \times 10^{11}$ cm$^{-2}$), and (2) the very low
and high density regimes where the 2-impurity model is no longer
operational are completely out of the experimentally relevant density
range of interest in high-mobility 2D semiconductor structures for the
physics (e.g., FQHE) studied in this context. Assuming a high-mobility
modulation doped GaAs quantum well of thickness $200 - 400$ \AA \; and
a set-back distance of $600 - 2000$ \AA \; (these are typical numbers
for high-mobility 2D GaAs structures), the 2-impurity model should be
well-valid in a wide range of carrier density $5\times 10^9$ cm$^{-2}
\alt n \alt 5\times 10^{11}$ cm$^{-2}$, which is the applicable
experimental regime of interest. Thus, the applicability of the
2-impurity model for considering the mobility/quality dichotomy is not
a serious issue of concern.

A second concern could be the validity (or not) of our theoretical
approximation scheme for calculating $\tau_{t,q}$, where we have used
the zero-temperature Boltzmann theory and the leading-order Born
approximation for obtaining the scattering rates. The $T=0$
approximation is excellent as long as $T \ll T_F = E_F/k_B$, which is
valid in all systems of interest in this context. For high-mobility 2D
semiconductor structures of interest in the current work, where the
issue of the dichotomy of mobility/quality is relevant (since in
low-mobility samples, typically $\tau_t \approx \tau_q$), the leading
order theory (in the disorder strength) employed in our approximation
scheme should, however, be excellent since the conditions $n\gg n_B$
and $n \gg n_Re^{-2k_Fd_R}$ are both satisfied making Born
approximation essentially an exact theory in this manifestly very weak
disorder situation (consistent with the high carrier mobility under
consideration). An equivalent way of asserting the validity of Born
approximation in our theory is to note that the conditions $E_F \gg
\Gamma$ and $k_F l \gg 1$ are always satisfied in the regime of our
interest (with $\Gamma$ and $l$ being the quantum level broadening and
the transport mean free path respectively). A related issue, which is
theoretically somewhat untractable, is the possible effect of impurity
correlation effects \cite{correlation} on the mobility versus quality
question in 2D semiconductor structures. It is straightforward to
include impurity correlation effects among the dopant ions in our
transport theory, but unfortunately no sample-dependent experimental
information is available on impurity correlations for carrying out
meaningful theoretical calculations. We have carried out some
representative numerical calculations assuming model inter-impurity
correlations among the remote dopants, finding that such correlations
enhance both $\tau_{t}$ and $\tau_q$, as expected (with $\tau_q$ being
enhanced more than $\tau_t$ in general), compared with the completely
random impurity configuration results presented in the current
article, but our qualitative conclusions about the mobility/quality
dichotomy remain unaffected since the fact that $\tau_t$ and $\tau_q$ are
controlled respectively by background and remote scattering in
high-mobility modulation-doped structures continues to apply in the
presence of impurity correlation effects. We therefore believe that our current
theory involving Born approximation assuming weak leading order
disorder scattering from random uncorrelated quenched charged
impurities in the environment (both near and far) is valid in the
parameter regime of our interest.

Finally, we comment on the possibility of future experimental work to
directly verify (or falsify) our theory. Throughout this paper, we
have, of course, made extensive contact with the existing experimental
results which, in fact, have motivated our current theoretical work on
the mobility versus quality dichotomy. For a direct future
experimental test of the theory, it will be necessary to produce a
large number of high-mobility 2D semiconductor structures with
different fixed carrier densities and with varying values of the
remote dopant setback distance, and then measure the values of
$\tau_{t,q}$ in a large set of samples which are all characterized by
their high-density mobility. The measurement of the transport
relaxation time $\tau_t$ is simple since it is directly connected to
the carrier mobility $\mu$ (or conductivity $\sigma$): $\tau_t = m
\sigma/ne^2 = m\mu/e$. The measurement of the single-particle
relaxation time (or the quantum scattering time) $\tau_q$ is, however,
not necessarily trivial although its theoretical definition is very
simple. In particular, the Dingle temperature or equivalently the
Dingle level broadening $\Gamma_D$ obtained from the measured
temperature dependence of the amplitude of the 2D SdH oscillations may
not necessarily give the zero-field quantum scattering time $\tau_q$
defining the sample quality in our theoretical considerations
(i.e. $\Gamma_D = \hbar/2\tau_q$ may not necessarily apply to the 2D
SdH measurements) because of complications arising from the quantum
Hall effect and inherent spatial density inhomogeneities (associated
with MBE growth) in the 2D sample. Since our theory is explicitly a
zero-magnetic field theory, it is more appropriate to obtain $\tau_q$
simply by carefully monitoring low-field magneto-resistance
oscillations finding the minimum magnetic field $B_0$ where the
oscillations disappear. The corresponding cyclotron energy $\omega_0
= e B_0/mc$ then defines the single particle level-broadening $\Gamma
\sim \hbar \omega_0$, providing $\tau_q = 1/2\omega_0$. An advantage
of this method of obtaining $\tau_q$ is that one is necessarily
restricted to the low magnetic field regime in high-mobility systems
(i.e., $E_F \gg \hbar \omega_0$), where our theory should be
applicable. A much stronger advantage be applicable. A much stronger
advantage of using this proposed definition (i.e., the disappearance
of magneto-resistance oscillations at the lowest experimental
temperature) for the experimental determination of $\tau_q$ is that
this is much easier to implement in the laboratory than the full
measurement of the Dingle temperature which requires accurate
measurements of the temperature dependent SdH amplitude
oscillations. We therefore suggest low-temperature measurements of
$\mu$ and $\omega_0$ to obtain $\tau_t$ and $\tau_q$ respectively in a
large number of modulation doped samples with varying $n$, $n_R$,
$d_R$, and $n_B$ in order to carry out a quantitative test of our
theory. A large systematic data base of both $\tau_t$ and $\tau_q$ in
many different samples should manifest poor correlations between these
two scattering times (i.e., the mobility/quality dichotomy) provided
the samples are high-mobility samples dominated by long-range charged
impurity disorder. As emphasized (and as is well-known) throughout this
work, if one type of disorder completely dominates both $\tau_t$ and
$\tau_q$, then they will obviously be correlated, but this should be
more an exception than the rule in high-mobility modulation-doped 2D
semiconductor structures, where both near (``$n_B$'') and far
(``$n_R$") impurities should, in general, play important roles in
manifesting the mobility/quality dichotomy.

One important open question is whether the mobility/quality dichotomy
we establish in the current work can be extended to other definitions
of quality beyond our definition of quality in terms of the
single-particle scattering rate or quantum level-broadening. The
advantage of using the quantum scattering rate as the measure of
sample quality is that this definition is generic, universal, and
simple to calculate (and to measure). However, obviously, given an
arbitrary disorder 
distribution involving long-range Coulomb-disorder, there are many
possible definitions of quality involving many different moments of
the effective Coulomb disorder. It will be very interesting for future
work to choose alternate possible definitions of sample quality to
establish whether our finding of the mobility/quality dichotomy
applies to all possible definitions of sample quality. 

This work is supported by LPS-CMTC, IARPA-ARO, Microsoft Q, and 
Basic Science
Research Program through the National Research Foundation of Korea
Grant funded by the Ministry of Science, ICT \& Future Planning
(2009-0083540).


\begin{thebibliography}{999}

\bibitem{stormer} H. L. St\"{o}rmer, R. Dingle, A. C. Gossard, and
  Wiegmann, Inst. Conf. Ser. London {\bf 43}, 557 (1978).

\bibitem{heiblum} J. P. Eisenstein, K. B. Cooper, L. N. Pfeiffer, and K. W. West, Phys. Rev. Lett. {\bf 88}, 076801 (2002);
M. Dolev, M. Heiblum, V. Umansky, A. Stern, and D.  Mahalu,
  Nature {\bf 452}, 829 (2008);   
W. Pan, J. S. Xia, H. L. Stormer, D. C. Tsui, C. Vicente, E. D. Adams, N. S. Sullivan, L. N. Pfeiffer, K. W. Baldwin, and K. W. West, Phys. Rev. B {\bf 77}, 075307 (2008);  
W. E. Chickering, J. P. Eisenstein, L. N. Pfeiffer, and K. W. West,
Phys. Rev. B 87, 075302 (2013); 
N. Deng, A. Kumar, M. J. Manfra, L. N. Pfeiffer, K. W. West, and G. A. Cs\'{a}thy, 
Phys. Rev. Lett. 108, 086803 (2012); W. Pan, J. S. Xia, H. L. Stormer, D. C. Tsui, C. Vicente,
  E. D. Adams, N. S. Sullivan, L. N. Pfeiffer, K. W. Baldwin, and
  K. W. West,  Phys. Rev. B {\bf 77}, 075307 (2008);
Yanhua Dai, R. R. Du, L. N. Pfeiffer, and K. W. West,
Phys. Rev. Lett. {\bf 105}, 246802 (2010);
V. Umansky, M. Heiblum, Y. Levinson, J. Smet, J. N\"{u}bler, M. Dolev, J. Cryst. Growth {\bf 311}, 1658 (2009); M. Manfra, arXiv:1309.2717;  L. Pfeiffer and K.W West, Physica E {\bf 20}, 57 (2003).


\bibitem{tsuiprl1982} D. C. Tsui, H. L. St\"{o}rmer, and A. C. Gossard, Phys. Rev. Lett. {\bf 48}, 1559 (1982).

\bibitem{willettprl1987}
R. Willett, J. P. Eisenstein, H. L. St\"{o}rmer, D. C. Tsui, A. C. Gossard, and J. H. English, Phys. Rev. Lett. {\bf 59}, 1776 (1987)

\bibitem{shayeganprl1992}
Y. W. Suen, L. W. Engel, M. B. Santos, M. Shayegan, and D. C. Tsui, Phys. Rev. Lett. {\bf 68}, 1379 (1992); J. P. Eisenstein, G. S. Boebinger, L. N. Pfeiffer, K. W. West, and Song He,
Phys. Rev. Lett. {\bf 68}, 1383 (1992).

\bibitem{lillyprl} M. P. Lilly,  K. B. Cooper, J. P. Eisenstein, L. N. Pfeiffer, and K. W. West, Phys. Rev. Lett. {\bf 82}, 394 (1999).

\bibitem{stormerbg} H. L. Stormer, L. N. Pfeiffer, K. W. Baldwin, and K. W. West,
Phys. Rev. B {\bf 41}, 1278(R) (1990);
 T. Kawamura and S. Das Sarma, \prb \; {\bf 45}, 3612
  (1992); T. Kawamura and S. Das Sarma, \prb \; {\bf 42}, 3725 (1990);
 H. Min, E. H. Hwang, and S. Das Sarma, Phys. Rev. B {\bf 86}, 085307
 (2012). 


\bibitem{dassarmaprb2013} S. Das Sarma and E. H. Hwang, Phys. Rev. B {\bf 88}, 035439 (2013);
E. H. Hwang and S. Das Sarma,  Phys. Rev. B {\bf 77}, 235437 (2008).

\bibitem{andormp} T. Ando, A. B. Fowler, and F. Stern,
  Rev. Mod. Phys. {\bf 54},  437 (1982). 

\bibitem{dassarmaPRB1985} S. Das Sarma and F. Stern, Phys. Rev. B {\bf 32}, 8442 (1985).

\bibitem{exp11} J. P. Harrang, R. J. Higgins, R. K. Goodall, P. R. Jay, M. Laviron, and P. Delescluse, Phys. Rev. B {\bf 32}, 8126 (1985); R. G. Mani and J. R. Anderson, {\it ibid}. {\bf 37}, 4299 (1988); M. Sakowicz, J. Lusakowski, K. Karpierz, M. Grynberg, and B. Majkusiak, Appl. Phys. Lett. {\bf 90}, 172104 (2007);
B. Das, S. Subramaniam, M. R. Melloch, and D. C. Miller,
Phys. Rev. B {\bf 47}, 9650 (1993);
E K Pettersen, D A Williams, and H Ahmed, Semicond. Sci. Technol. {\bf 11}, 1151 (1996);
P. T. Coleridge, R. Stoner, and R. Fletcher,
Phys. Rev. B {\bf 39}, 1120 (1989);
P. T. Coleridge, Phys. Rev. B {\bf 44}, 3793 (1991).

\bibitem{gamez} G. Gamez and K. Muraki, Phys. Rev. B {\bf 88}, 075308 (2013).


\bibitem{IRC} S. Das Sarma and E. H. Hwang, arXiv:1401.4762.

\bibitem{review} E. Abrahams, S. V. Kravchenko, and M. P. Sarachik,
  Rev. Mod. Phys. {\bf 73}, 251 (2001); S. V. Kravchenko, and M. P. Sarachik, Rep. Prog. Phys. {\bf 67}, 1 (2004); S. Das Sarma and E. H. Hwang, Solid State Commun. {\bf 135}, 579 (2005);
B. Spivak, S. V. Kravchenko, S. A. Kivelson, and X. P. A. Gao, Rev. Mod. Phys. {\bf 82}, 1743 (2010);
S. Das Sarma, S. Adam, E. H. Hwang, and E. Rossi, Rev. Mod. Phys. {\bf 83}, 407 (2011).

\bibitem{lilly} M. P. Lilly, J. L. Reno, J. A. Simmons, I. B. Spielman, J.
P. Eisenstein, L. N. Pfeiffer, K. W. West, E. H. Hwang,
and S. Das Sarma, Phys. Rev. Lett. {\bf 90}, 056806 (2003);
S. Das Sarma, M. P. Lilly, E. H. Hwang, L. N. Pfeiffer,
K. W. West, and J. L. Reno, Phys. Rev. Lett. {\bf 94}, 136401 (2005);

\bibitem{panprl2011} W. Pan, N. Masuhara, N. S. Sullivan, K. W. Baldwin, K. W. West, L. N. Pfeiffer, and D. C. Tsui, Phys. Rev. Lett. {\bf 106}, 206806 (2011).


\bibitem{samkharadze} N. Samkharadze, J. D. Watson, G. Gardner, M. J. Manfra, L. N. Pfeiffer, K. W. West, and G. A. Cs\'{a}thy, Phys. Rev. B 84, 121305 (2011).


\bibitem{choi}   H. C. Choi, W. Kang, S. Das Sarma,
  L. N. Pfeiffer, and K. W. West,  Phys. Rev. B {\bf 77}, 081301(R)
  (2008); J. S. Xia, W. Pan, C. L. Vicente, E. D. Adams, N. S. Sullivan, H. L. Stormer, D. C. Tsui, L. N. Pfeiffer, K. W. Baldwin, and K. W. West, Phys. Rev. Lett. {\bf 93}, 176809 (2004).


\bibitem{dean} C. R. Dean, B. A. Piot, P. Hayden, S. Das Sarma, G. Gervais, L. N. Pfeiffer, and K. W. West, Phys. Rev. Lett. {\bf 100}, 146803 (2008).

\bibitem{kane} B. E. Kane, L. N. Pfeiffer, K. W. West, and C. K. Harnett, Appl. Phys. Lett. {\bf 63}, 2132 (1993).

\bibitem{duprl1993}R. R. Du, H. L. Stormer, D. C. Tsui, L. N. Pfeiffer, and K. W. West
Phys. Rev. Lett. {\bf 70}, 2944 (1993).

\bibitem{duprivate} R. R. Du, private communications.


\bibitem{morf} M. Storni, R. H. Morf, and S. Das Sarma,
Phys. Rev. Lett. {\bf 104}, 076803 (2010).

\bibitem{morf2} R. H. Morf, N. d'Ambrumenil, and S. Das Sarma, Phys. Rev. B {\bf 66}, 075408 (2002).


\bibitem{sajorto} T. Sajoto, Y. W. Suen, L. W. Engel, M. B. Santos, and M. Shayegan,
Phys. Rev. B {\bf 41}, 8449 (1990). 

\bibitem{correlation} T. Kawamura and S. Das Sarma, Solid State Commun. {\bf 100}, 411 (1996); 
S. Das Sarma and S. Kodiyalam, Semicond. Sci. Technol. {\bf 13}, A59 (1998);
Qiuzi Li, E. H. Hwang, E. Rossi, and S. Das Salma,
Phys. Rev. Lett. {\bf 107}, 156601 (2011); E. Buks, M. Heiblum, and Hadas Shtrikman,
Phys. Rev. B {\bf 49}, 14790(R) (1994); 
A. F. J. Levi, S. L. McCall and P. M. Platzman, Appl. Phys. Lett. {\bf 54}, 940 (1989).




\end{thebibliography}
\end{document}